\documentclass{article}

\usepackage{arxiv}

\usepackage[utf8]{inputenc} 
\usepackage[T1]{fontenc}    
\usepackage{hyperref}       
\usepackage{url}            
\usepackage{booktabs}       
\usepackage{amsfonts}       
\usepackage{nicefrac}       
\usepackage{microtype}      
\usepackage{lipsum}
\usepackage{graphicx}
\usepackage{placeins}
\usepackage{xcolor}
\usepackage{verbatim}
\graphicspath{ {./images/} }

\usepackage{listings}

\lstdefinelanguage{mlir}{
  keywords =
  {addf,affine,alloc,apply,call,memref,dealloc,dma_start,dma_wait,if,for,memref,mulf,return,splat,tensor,load,store},
	keywordstyle=\color{blue}\bfseries,
	ndkeywordstyle=\color{darkgray}\bfseries,
  identifierstyle=\color{black},
	sensitive=false,
	comment=[l]{//},
  commentstyle=\color{purple}\ttfamily,
}

\lstdefinelanguage{Shell}{
	keywords = {},
	keywordstyle=\color{blue}\bfseries,
	ndkeywordstyle=\color{darkgray}\bfseries,
  identifierstyle=\color{black},
	sensitive=false,
	comment=[l]{//},
	morecomment=[s]{/*}{*/},
  commentstyle=\color{purple}\ttfamily,
	stringstyle=\color{red}\ttfamily,
	morestring=[b]',
	morestring=[b]"
}

\lstdefinestyle{cmd} {
	language=Shell,
  columns=flexible,
  basicstyle=\small\ttfamily,
	literate={\$}{{\textcolor{blue}{\$}}}1
}

\lstdefinestyle{mlir} {
	language=mlir,
  basicstyle=\small\ttfamily,
  columns=flexible,
  mathescape
}

\usepackage{xifthen}

\title{High Performance Code Generation in MLIR: An Early Case Study with GEMM}

\author{
  Uday Bondhugula \\
  Dept of CSA, Indian Institute of Science \& PolyMage Labs \\
  Bengaluru, Karnataka, India. \\
  \texttt{udayb@iisc.ac.in, uday@polymagelabs.com} \\
}

\begin{document}

\date{}

\maketitle

\begin{abstract}

This article is primarily meant to present an early case study on using
MLIR~\cite{mlir2020arxiv}, a new compiler intermediate representation
infrastructure, for high-performance code generation. Aspects of MLIR covered
in particular include memrefs, the affine dialect, and polyhedral utilities and
pass infrastructure surrounding those.  This article is also aimed at showing
the role compiler infrastructure could play in generating code that is
competitive with highly tuned manually developed libraries, albeit in a more
modular, reusable, and automatable way.

\end{abstract}


\section{Introduction, Motivation, and Setup}

It is well-known that the computationally demanding routines driving the
state-of-the-art in domains such as dense linear algebra and deep learning are
all based on carefully hand-optimized and tuned libraries developed with
significant engineering effort and insights over time. The techniques and the
level of attention necessary to optimize for near-peak performance on a target
architecture makes the process often inaccessible to those without a deep
knowledge of code optimization and low-level interactions with hardware.  In
many cases, the best performing code comes from the hardware vendors themselves.
Yet, fortunately, there are published works such as those of Goto and Van de
Geijn~\cite{goto2008toms} that have described in great detail how such close to
peak performance could be obtained.  Subsequent works~\cite{vanzee2015toms}
made the process more modular, reusable, and accessible to a wider audience,
having translated to an open-source project
\href{https://github.com/flame/blis}{FLAME/BLIS}.

This article alludes to the fact that this process could potentially be made
even more modular, automatable and systematic --- by hosting it on top of a real
compiler IR that has the necessary abstractions and infrastructure. This
completely avoids the need to write any code by hand --- be it C or inline
assembly. The IR infrastructure that will be used here is of course,
MLIR~\cite{mlir2020arxiv,mlir-web} and we will try to recreate the
OpenBLAS/BLIS' optimization approach in a compiler-oriented manner using MLIR.
MLIR is a new intermediate representation designed to provide a unified,
modular, and extensible infrastructure to progressively lower dataflow compute
graphs, through loop nests potentially, to high-performance target-specific
code. While MLIR shares similarities with traditional CFG-based three-address
SSA representations (including LLVM IR and Swift intermediate language), it also
introduces notions from the polyhedral compiler framework as first class
concepts to allow powerful analysis and transformation in the presence of loop
nests and multi-dimensional arrays. It is thus a useful infrastructure suitable
for developing new compilers, whether general-purpose or domain-specific, and
whether targeting general-purpose multicores or specialized accelerator chips.

\subsection{Setup}

We are going to be using an Intel Skylake-based high-end desktop/workstation
processor for all experimentation. The processor is an \href{https://ark.intel.com/content/www/us/en/ark/products/126684/intel-core-i7-8700k-processor-12m-cache-up-to-4-70-ghz.html}{Intel(R) Core(TM)
i7-8700K CPU @ 3.70GHz}, which is actually based on
\href{https://en.wikichip.org/wiki/intel/microarchitectures/coffee_lake}{Coffee
Lake}, a process refinement of Skylake, and thus with the same core/performance
characteristics.  Note
that although this is based on the Skylake microarchitecture, it is not a
SkyLake-X: as such, its vectors are not AVX-512 but just AVX-2 (256-bit). It has
a 32 KiB L1 data cache and a 256 KiB L2 unified cache per core, and a 12 MiB
shared L3 cache (2 MiB / core).

Before we start, we are going to compute the machine peak on this processor.
This CPU supports AVX-2 and has two FMA units that can operate on 256-bit wide
vectors.  As such, one can perform $2 * 4 * 2$ double-precision floating-point
multiply and adds per cycle, which at the max turbo boost frequency of 4.7~GHz
translates to 75.2~GFLOPS per core. (Frequency scaling can impact how this
figure is calculated, but we will ignore that for now.)

\subsection{Running example: DGEMM}

Matrix-matrix multiplication is often an excellent routine for a tutorial or
exercise in code optimization because a number of standard practices could be
demonstrated using it. It is also one of the most important ones for many
domains and thus often the first thing one builds an optimized implementation
for when rolling out a new architecture.

\subsection{Starters}

Let us take a 2088x2048 double precision matrix-matrix multiplication, and
benchmark it in various ways. We choose 2088 here since it is divisible by both
4 and 3: this makes it easier to read some of the snippets given that we will be
exploring register tiling by sizes that are multiples either of 2, 3, or 4.

\subsubsection{Compilers with -O3, Pluto}

Let us see what current compilers do if this is written as a naive nest in C.
These results can be reproduced with the setup in
\href{https://github.com/bondhugula/pluto}{Pluto} from under examples/matmul/. (We
use -ffast-math along with -O3 to be fair given the other things we are going
to compare this with.)

\begin{lstlisting}[style=cmd]
$ clang -v
clang version 8.0.0 (Fedora 8.0.0-3.fc30)
Target: x86_64-unknown-linux-gnu

$ clang -O3 -ffast-math -DTIME matmul.c -o matmul.clang -lm
$ ./matmul.clang
36.484855s
0.47 GFLOPS

$ gcc --version
gcc (GCC) 9.2.1 20190827 (Red Hat 9.2.1-1)

$ gcc -O3 -ffast-math  -DTIME matmul.c -o matmul.gcc -lm
$ ./matmul.gcc
3.788421s
4.53 GFLOPS
\end{lstlisting}

Disappointingly, clang and GCC are at 0.6\% and 6\% of the machine peak
respectively! But general-purpose compilers are not expected or meant to get
anywhere to the peak. Programmers instead typically use highly tuned libraries.
We are going to get to that shortly, but a quick note on why Clang performs
poorly here.  Clang only performs innermost loop vectorization, which is really
a bad choice here (due to poor locality) when that's the only thing being done.
Even a textbook loop interchange here from $ijk \rightarrow ikj$ improves Clang
performance by about 10x (because now the right loop would be at the innermost
level for both locality and auto-vectorization). Clang does not perform any
unroll-and-jam here either. While on this let us also see what a polyhedral tool
\href{https://github.com/bondhugula/pluto}{Pluto}, a source to source
translator, does on this. The below run shows that it is at about 25\% of the
machine peak, which is better but also pretty unsatisfying.

\begin{lstlisting}[style=cmd]
$ make tiled
$ ./tiled
0.889177s
19.32 GFLOPS
\end{lstlisting}

\subsubsection{Libraries}

Now, let us see what the fastest libraries in the world do on matmul. Again, the
setup in \href{https://github.com/bondhugula/pluto/tree/master/examples/matmul}{Pluto} allows
experimenting with OpenBLAS, BLIS, or MKL quickly.

\begin{lstlisting}[style=cmd]
$ make mkl blis openblas

$ ./blis
63.29 GFLOPS

$ ./openblas
0.259527s
67.49 GFLOPS

$ ./mkl
0.273492s
67.70 GFLOPS
\end{lstlisting}

MKL and OpenBLAS are nearly at 92\% of the peak (68/75.2), while BLIS is close at
85\% of the peak. We will keep these in mind as we try out how to build
high-performance versions with MLIR. Note: we haven't been very rigorous with
the above experimentation in terms of performing warmup runs, taking an average
across repetitions etc., but our goal is to just get a reasonably accurate
picture for now.

\section{Using MLIR to Optimize GEMM}

There is currently no C/C++ or another language/programming model frontend that
emits MLIR. The way to get something in MLIR run on CPUs is through {\it
mlir-cpu-runner} which can take MLIR as input and JIT compile and execute it.
As part of this process, optimized
MLIR is converted to LLVM IR, which then goes through its optimization pipeline
to generate target code, all of this through {\it mlir-cpu-runner}.

\subsection{Affine Ops}
Here is how a simple canonical matrix-matrix multiplication looks like in MLIR
-- the structure is close to its form in C.

\begin{lstlisting}[style=mlir]
// C += A * B.
func @matmul(%A: memref<2048x2048xf64>, %B: memref<2048x2048xf64>, %C: memref<2048x2048xf64>) {
  affine.for %arg3 = 0 to 2048 {
    affine.for %arg4 = 0 to 2048 {
      affine.for %arg5 = 0 to 2048 {
        %a = affine.load %A[%arg3, %arg5] : memref<2048x2048xf64>
        %b = affine.load %B[%arg5, %arg4] : memref<2048x2048xf64>
        %ci = affine.load %C[%arg3, %arg4] : memref<2048x2048xf64>
        %p = mulf %a, %b : f64
        %co = addf %ci, %p : f64
        affine.store %co, %C[%arg3, %arg4] : memref<2048x2048xf64>
      }
    }
  }
  return
}

func @main() {
  %A = alloc() : memref<2048x2048xf64>
  %B = alloc() : memref<2048x2048xf64>
  %C = alloc() : memref<2048x2048xf64>

  %cf1 = constant 1.00000e+00 : f64

  linalg.fill(%A, %cf1) : memref<2048x2048xf64>, f64
  linalg.fill(%B, %cf1) : memref<2048x2048xf64>, f64
  linalg.fill(%C, %cf1) : memref<2048x2048xf64>, f64

  call @matmul(%A, %B, %C) : (memref<2048x2048xf64>, memref<2048x2048xf64>, memref<2048x2048xf64>) -> ()
  call @print_memref_2d_f64(%C): (memref<2048x2048xf64>) -> ()
  return
}

func @print_memref_2d_f64(memref<2048x2048xf64>)
\end{lstlisting}

affine.for and affine.load/store are ops from the Affine
dialect~\cite{affine-dialect} used above. The IR above also uses a helper op
from the LinAlg dialect to initialize matrices. The rest like (addf, mulf,
alloc) are all ops from MLIR's standard dialect.

\subsection{MemRefs}

A memref~\cite{memref} in MLIR is the in-memory representation of a
tensor in MLIR.  Depending on its layout
map, it could refer to a potentially non-contiguous
sub-region of the underlying buffer. All memrefs in the above snippet have the
default identity layout map $(d0, d1) \rightarrow (d0, d1)$, which corresponds
to a row major layout in memory.  2088 x 2048 is the shape of memref where each
of its elements is an f64 (double). There is other information such as an
affine layout map (a mapping for its logical coordinate space to physical
memory) that is elided when these are identity maps (the default). We will look
at an example of this a little later (Section~\ref{sec:affine-map-detour}).
The only way to access the elements of a memref is through load and store
operations.

\subsection{A high-level domain-specific op that can expand}

MLIR is extensible in several ways, and one of them is in the ease of
adding/creating new ops at the level of abstraction we desire. One way of
classifying ops or IR in general in MLIR is into high-level, mid-level, and
low-level, although there is not a clear distinction and lowering can be
progressive and you can have a mix. High-level ops operate on tensor and memref
types themselves, i.e., their inputs (operands) and results are of that type.
affine.for and affine.if are mid-level ops -- these have nested blocks (which
are themselves a list of ops) and a particular structure to them.  Low-level
ops are at the same level as instructions in LLVM, and these typically
correspond to either one or a flat sequence of instructions matching a target
(three address code style in other words).

If one is building a DSL and knows that they need or have a matmul, one could
just define an op that does the matmul taking memrefs as operands. For the
purpose of this article, we will create a hop.matmul op.  An operation in MLIR
has inputs, results, and attributes (ones with regions have region arguments as
well).

\begin{lstlisting}[style=mlir]
hop.matmul %A, %B, %C {some_attr = 96, another_attr = 4}
       : (memref<2088x2048xf64>, memref<2048x2048xf64>, memref<2088x2048xf64>)
\end{lstlisting}

hop.matmul above is an operation that operates on three memrefs, \%A, \%B, and
\%C, which are the LHS, the RHS, and the output corresponding to a
matrix-matrix multiplication. This op has zero results: since these are
memrefs, the underlying memory can be updated, but the memref itself is an SSA
value and thus can be defined only once. In this case, \%C will be both read
and written by this heavy op.  {\it some\_attr} is an attribute of the op,
which like with LLVM metadata is meant to be a constant of one of the numerous
types available including some polyhedral types like
\href{https://github.com/llvm/llvm-project/tree/master/mlir/docs/Dialects/Affine.md#affine-maps}{affine
maps and integer sets}.  Such polyhedral attributes could be used to encode
powerful information and we will see an example of this later.

For the purpose of this article, we create an -hopt pass that is able to expand
out hop.matmul into the naive 3-d loop nest we showed further above. Let us just
execute this with MLIR without any optimization to see where we are starting our
journey from.

\begin{lstlisting}[style=cmd]
// This is where we start. No optimizations performed in MLIR besides canonicalization.
$ mlir-opt -hopt -convert-linalg-to-loops -lower-affine -convert-std-to-llvm dgemm.mlir |
mlir-cpu-runner -O3 -e main -entry-point-result=void -shared-libs=lib/libmlir_runner_utils.so
Compilation time: 0.015995s
0.558177 GFLOPS
\end{lstlisting}

This is pretty close to the execution time of `clang -O3' that we saw above,
which makes sense, but this is equally slow (less than 1\% of the machine peak)!
But we have not really made use of or run any of the optimization infrastructure
in MLIR. The -hopt here just lowers this op into the canonical 3-d loop nest we
showed further above, which is then lowered to LLVM and goes through the same
LLVM -O3 pipeline that clang went through.

\subsection{Tiling for caches: OpenBLAS/BLIS tiling as a polyhedral schedule
}

Tiling and blocking has been studied for several decades, and it is now known
how matrix-matrix multiplication should be tiled for high performance on widely
used architectures. We are going to use the same cache tiling strategy used in
OpenBLAS and BLIS. [This scheme](https://dl.acm.org/citation.cfm?id=1356053) is
a very clever one that carefully tiles for multiple levels of the cache in a way
that exploits reuse for each of the matrices involved at one level or another:
the objective is to make sure the vector FMA/add mul pipeline remains full and
one does not wait for loads. It does not blindly tile the nest multiple times to
exploit reuse for all matrices at L1, L2, and L3, as would most polyhedral tools
do.

There are various ways of describing OpenBLAS's/BLIS' tiling scheme and they are
explained in excellent detail with illustration in the papers by
\href{https://dl.acm.org/citation.cfm?id=2764454}{Van Zee and
van de Geijn on BLIS}~\cite{vanzee2015toms}, and later
with analysis and modeling by
\href{https://dl.acm.org/citation.cfm?id=2764454}{Low et al.
2016}~\cite{low2016toms}.  But here, we will describe
the tiling transformation compactly via polyhedral schedules. For an
introduction to polyhedral schedules, see
\href{https://www.csa.iisc.ac.in/~udayb/slides/uday-polyhedral-opt.pdf}{here}~\cite{uday-poly-intro}
or the
material on \href{https://polyhedral.info}{polyhedral.info}. In particular, the
tiling and corresponding intra-tile loop orders used in the OpenBLAS/BLIS
approach can be expressed as the following polyhedral schedule (a
multi-dimensional affine map / transformation):
\[
(i, j, k) \rightarrow (j / N_C, k / K_C, i / M_C, j / N_R, i / M_R, k \ \textrm{ mod } \ K_C, j
\ \textrm{ mod } \ N_R, i \textrm{ mod } M_R),
\]
where `/' is a floor division. The innermost two loops after applying the above
schedule are meant to be fully unrolled leading to an $M_R$ x $N_R$ loop body.
Ignoring the last level of tiling (for the L3 cache), this becomes:
\[
(i, j, k) \rightarrow (k / K_C, i / M_C, j / N_R, i / M_R, k \textrm{ mod } K_C, j \textrm{ mod } N_R,
i \textrm{ mod } M_R).
\]
Figure~\ref{fig:blis-tiling} from the BLIS works is a great illustration of its
optimization strategy:
\begin{figure}[!h]
	\centering
  \includegraphics[width=0.75\linewidth]{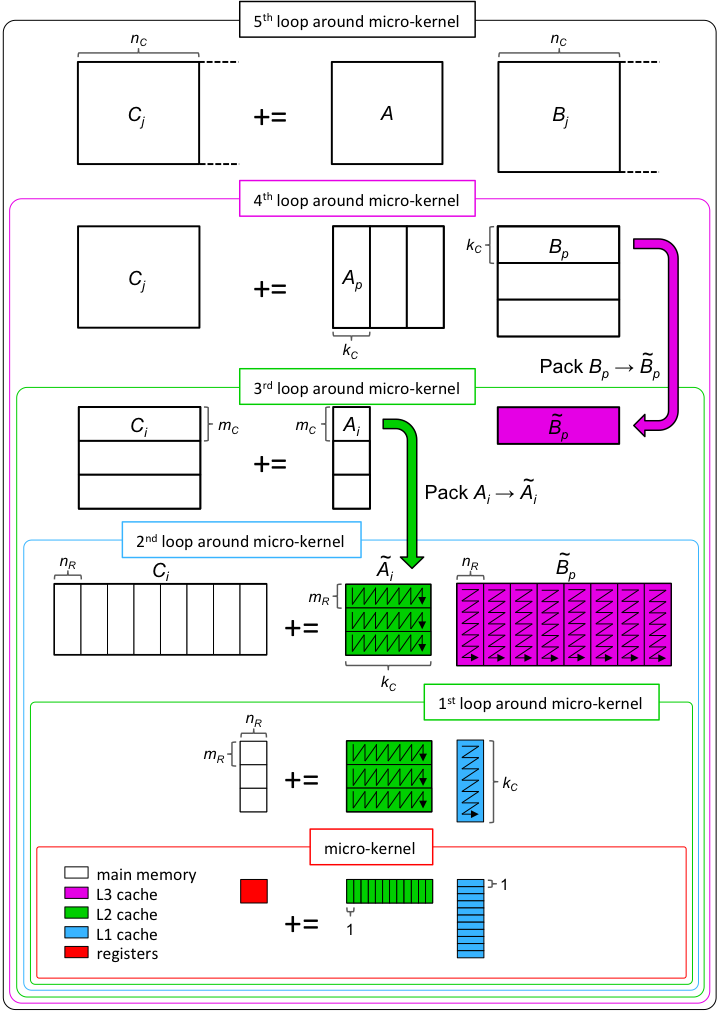}
  \caption{
Tiling strategy of OpenBLAS/BLIS. Figure courtesy: Field Van
  Zee and Robert van de Geijn~\label{fig:blis-tiling}}
\end{figure}

The parameters ($M_C$, $K_C$) are chosen such that a tile of the LHS matrix
(\%A) of
size $M_C$ x $K_C$ is reused in the L2 cache, an RHS matrix tile of size $K_C$ x $N_R$
is reused in the L1 cache, while a register tile of the output matrix 
$M_R$ x $N_R$ is reused in just the registers. There are several other
considerations in choosing these parameters if one is interested in analytical
modeling; those are described in the work of Low et al.~\cite{low2016toms}.
The last three dimensions of
the schedule are multiplying a panel of \%A of size $M_R$ x $K_C$ with a panel of \%B
of size $K_C$ x $N_R$. Note that \%C is being both read and written, and
intuitively,
it makes sense to exploit reuse for it in the registers while running a large
enough k intra-tile loop at the innermost level. In the schedule above, the
innermost two dimensions (j, i) are fully unrolled, i.e., you get an unrolled
body of size $M_R$ x $N_R$ times the original one. For readers seeking more detail
in general on this and on programming for high performance, please see \href{http://www.cs.utexas.edu/users/flame/laff/pfhp/}{course notes}~\cite{flame-laff} from the BLIS team members.

\subsection{Tiling in MLIR}

There are two ways of achieving this tiling in MLIR: one by calling
[mlir::tile]
(which is also what the loop tiling pass in MLIR uses), and then performing the
desired loop interchange via
\href{https://github.com/llvm/llvm-project/tree/master/mlir/lib/Transforms/Utils/LoopUtils.cpp#L510}{mlir::interchangeLoops}
The other is by implementing a higher order polyhedral (HOP) approach based on
domains and schedules. We use this latter approach here since MLIR's affine
analysis machinery does not yet have the necessary simplification to get rid of
certain redundant bounds resulting from tiled code generation in advanced cases.
The HOP approach depends on an external library,
[ISL](http://isl.gforge.inria.fr), and we implement this as part of the -hopt
pass. We express the tiling schedule as an MLIR
\href{https://github.com/llvm/llvm-project/tree/master/mlir/docs/Dialects/Affine.md#affine-maps}{affine map}
(in fact, any affine schedule could be used), perform the code generation via
ISL, and convert the ISL AST back to MLIR. We will now use $M_C$, $K_C$, $M_R$, $N_R$ as
\href{https://github.com/llvm/llvm-project/tree/master/mlir/docs/LangRef.md#attributes}{attributes}.
on the hop.matmul op.

\begin{lstlisting}[style=mlir]
hop.matmul %A, %B, %C { M_C = 64 : i32, K_C = 256 : i32, M_R = 4 : i32, N_R = 8 : i32}
  : (memref<2088x2048xf64>, memref<2048x2048xf64>, memref<2048x2048xf64>)
\end{lstlisting}

We have used $K_C$ = 256, $M_C$ = 64, $M_R$ = 4, $N_R$ = 8 as a starting point: these can
be analyzed to be good values given the cache size constraints described
earlier. This is functionally equivalent to using the following schedule with
HOP and fully unrolling the last two dimensions, d1, d0, which will be of trip
counts 8 and 4 respectively):

\begin{lstlisting}[style=mlir]
hop.matmul %A, %B, %C {
    schedule = (d0, d1, d2) -> (d2 floordiv 256, d0 floordiv 64, d1 floordiv 8, d0 floordiv 4, d2, d1, d0)
  } : (memref<2088x2048xf64>, memref<2048x2048xf64>, memref<2088x2048xf64>)
\end{lstlisting}

In order to just evaluate cache tiling for now, we will just use the following schedule
(drop the tiling for registers):

\begin{lstlisting}[style=mlir]
hop.matmul %A, %B, %C {
    schedule = (d0, d1, d2) -> (d2 floordiv 256, d0 floordiv 64, d1, d0, d2)
  } : (memref<2088x2048xf64>, memref<2048x2048xf64>, memref<2088x2048xf64>)
\end{lstlisting}

\FloatBarrier

Now, with the above tiling, we get this loop nest:

\begin{lstlisting}[style=mlir]
...
#map8 = (d0) -> (d0 * 16)
#map9 = (d0) -> (522, d0 * 16 + 16)
...
func @matmul(%A: memref<2088x2048xf64>, %B: memref<2048x2048xf64>, %C: memref<2088x2048xf64>) {
  %c0 = constant 0 : index
  affine.for %arg3 = 0 to 8 {
    affine.for %arg4 = 0 to 33 {
      affine.for %arg5 = 0 to 2048 {
        affine.for %arg6 = #map7(%arg4) to min #map8(%arg4) {
          %0 = alloca() : memref<1xf64>
          %1 = affine.load %C[%arg6, %arg5] : memref<2088x2048xf64>
          affine.store %1, %0[%c0] : memref<1xf64>
          affine.for %arg7 = 0 to 256 {
            %3 = affine.load %A[%arg6, %arg3 * 256 + %arg7] : memref<2088x2048xf64>
            %4 = affine.load %B[%arg3 * 256 + %arg7, %arg5] : memref<2048x2048xf64>
            %5 = affine.load %0[0] : memref<1xf64>
            %6 = mulf %3, %4 : f64
            %7 = addf %5, %6 : f64
            affine.store %7, %0[0] : memref<1xf64>
          }
          %2 = affine.load %0[%c0] : memref<1xf64>
          affine.store %2, %C[%arg6, %arg5] : memref<2088x2048xf64>
        }
      }
    }
  }
  return
}
\end{lstlisting}

Note that the invariant load on \%B has been hoisted out. When we execute this:

\begin{lstlisting}[style=cmd]
// With tiling.
$ mlir-opt -hopt -convert-linalg-to-loops -lower-affine -convert-std-to-llvm dgemm.mlir |
mlir-cpu-runner -O3 -e main -entry-point-result=void -shared-libs=lib/libmlir_runner_utils.so
Compilation time: 0.0443082s
1.6012 GFLOPS
\end{lstlisting}

This is nearly a 3x improvement in performance, but still nowhere close to the
machine peak or the performance of MKL, OpenBLAS or BLIS!

\subsection{Explicit Copying or Packing}

Whenever a code with multi-dimensional arrays exhibits reuse, explicit
copying or packing is a commonly used technique in HPC, which first copies
or packs accessed data into contiguous buffers and indexes those buffers for
computation. Such copying reduces or nearly eliminates conflict misses, TLB
misses, and improves hardware prefetching performance (since the access will
lead to fewer prefetch streams). More importantly, in conjunction with tiling,
it provides the intended benefits of tiling. On the one hand, tiling is
performed so that reuse is exploited in multiple directions when the data
accessed fits in a higher level of the memory hierarchy, on the other hand, the
data accessed for a tile is no longer contiguous in the original matrix/tensor
leading to conflict misses, TLB misses, and more prefetch streams --- taking
away a lot of the gain even if there is high reuse. There is also a choice in
terms of which depth / where one would like to introduce the copies, but this
is quite clear for the tiling strategy we are using.

The packing optimization is something that one would really wish a compiler
performs automatically. MLIR has a pass
-affine-data-copy-generate
and a utility
{\it mlir::affineDataCopyGenerate}
that can perform explicit copying or packing.  The utility can also be called on
a specific memref to perform the packing at a specific loop depth, and with a
number of other options (including DMA support).

\begin{figure}[!ht]
  \centering
  \includegraphics[width=0.45\linewidth]{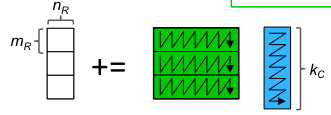}
  \caption{The one in green is the LHS to be packed into a buffer that will
  reside in L2, and the RHS in blue will reside in L1. (Figure courtesy: Field
  Van
  Zee and Robert van de Geijn). \label{fig:packing}}
\end{figure}

With the tiling approach we are using here, packing is performed for the LHS
matrix \%A right under the second loop ($i \textrm{ floordiv } K_C$ dimension in
the schedule). For \%B (RHS matrix), the copying is performed right under j
floordiv $N_R$ (the third dimension). Figure~\ref{fig:packing} illustrates the
packing needed.

\FloatBarrier

To perform the packing automatically with MLIR, instead of calling the
-affine-data-copy-generate pass, we will just use the underlying utility
[mlir::affineDataCopyGenerate] to place copies at the right depths.  For the
purpose of this article, we enabled this under the -hopt-copy option.  With
-hopt-copy, now the nest looks like:

\begin{lstlisting}[style=mlir]
affine.for %arg3 = 0 to 8 {
  affine.for %arg4 = 0 to 32 {
    %0 = alloc() : memref<64x256xf64>
    // Packing %A into a 64x256 buffer.
    affine.for %arg5 = #map6(%arg4) to #map7(%arg4) {
      affine.for %arg6 = #map4(%arg3) to #map5(%arg3) {
        %1 = affine.load %arg0[%arg5, %arg6] : memref<2048x2048xf64>
        affine.store %1, %0[%arg4 * -64 + %arg5, %arg3 * -256 + %arg6] : memref<64x256xf64>
      }
    }
    affine.for %arg5 = 0 to 256 {
      %1 = alloc() : memref<256x8xf64>
      // Packing %B into a 256x8 buffer.
      affine.for %arg6 = #map4(%arg3) to #map5(%arg3) {
        affine.for %arg7 = #map9(%arg5) to #map10(%arg5) {
          %2 = affine.load %arg1[%arg6, %arg7] : memref<2048x2048xf64>
          affine.store %2, %1[%arg3 * -256 + %arg6, %arg5 * -8 + %arg7] : memref<256x8xf64>
        }
      }
      affine.for %arg6 = #map16(%arg4) to #map17(%arg4) {
        // This is multiplying a packed 64x256 LHS panel with a packed 256x8 RHS panel.
        affine.for %arg7 = 0 to 256 {
          affine.for %arg8 = 0 to 8 {
            affine.for %arg9 = 0 to 4 {
              %2 = affine.load %0[%arg4 * -64 + %arg6 * 4 + %arg9, %arg7] : memref<64x256xf64>
              %3 = affine.load %1[%arg7, %arg8] : memref<256x8xf64>
              %4 = affine.load %arg2[%arg6 * 4 + %arg9, %arg5 * 8 + %arg8] : memref<2048x2048xf64>
              %5 = mulf %2, %3 : f64
              %6 = addf %4, %5 : f64
              affine.store %6, %arg2[%arg6 * 4 + %arg9, %arg5 * 8 + %arg8] : memref<2048x2048xf64>
            }
          }
        }
      }
      dealloc %1 : memref<256x8xf64>
    }
    dealloc %0 : memref<64x256xf64>
  }
}
\end{lstlisting}

\begin{lstlisting}[style=cmd]
// With tiling and packing.
$ mlir-opt -hopt -hopt-copy -convert-linalg-to-loops -lower-affine -convert-std-to-llvm
dgemm.mlir | mlir-cpu-runner -O3 -e main -entry-point-result=void
-shared-libs=lib/libmlir_runner_utils.so
Compilation time: 0.0506358s
2.38624 GFLOPS
\end{lstlisting}

This is nearly a 1.5x improvement in performance, but in absolute terms still
pretty bad. On a general note, the benefit of any optimization often depends on
how well optimized the code is on other aspects before the optimization was
applied -- even more so in this case. It's important to be mindful of this in
order to not draw incorrect conclusions on the relative benefits of each
individual transformation at this time. For example, the benefits of cache
tiling on a code where unroll-and-jam or register tiling has already been
performed is significantly lower than on a version where no optimizations were
performed. We will look at an example a little later.

\subsection{Unroll and Jam}

We will now use the complete schedule that has the $M_R$ and $N_R$ register tiling
so that an unroll-and-jam of the innermost two loops by $N_R$ and $M_R$
respectively could be achieved, but we actually didn't enable that yet. We use
the option -hopt-unroll to turn this on and off. We also run scalar replacement
in MLIR post unroll-and-jam. Besides turning memref locations being reduced
into scalars (single element memrefs) and hoisting them out, it also eliminates
redundant loads, and hoists invariant loads out of loops.

While at unroll-and-jam, we will also add one more parameter to the list,  $K_U$,
which will be the unroll factor for the $K_C$ loop (intra-tile loop corresponding
to k reduction dimension). We use $K_U$ = 4 to unroll the k loop (which will end
up as the innermost loop post all register tiling). So, we have:

\begin{lstlisting}[style=mlir]
hop.matmul %A, %B, %C { M_C = 64 : i32, K_C = 256 : i32, M_R = 4 : i32, N_R = 8 : i32, K_U = 4 : i32}
  : (memref<2088x2048xf64>, memref<2048x2048xf64>, memref<2088x2048xf64>)
\end{lstlisting}

\begin{lstlisting}[style=cmd]
// With tiling, packing, and unroll-and-jam/unroll.
$ mlir-opt -hopt -hopt-copy -hopt-unroll -hopt-scalrep -convert-linalg-to-loops
-lower-affine -convert-std-to-llvm dgemm.mlir | mlir-cpu-runner -O3 -e main
-entry-point-result=void -shared-libs=lib/libmlir_runner_utils.so
Compilation time: 0.11306s
23.2737 GFLOPS
\end{lstlisting}

A code is only as fast as its weakest link. While the cache tiling really
optimized reuse for the LHS and RHS, the strategy used by OpenBLAS and BLIS
exploits reuse for the output matrix only in registers (not in L1 nor L2).  As
such, without unroll-and-jam aka register tiling, brakes had been applied.  So,
this last step opened the floodgates here yielding a 10x improvement and
getting us to 23 GFLOPS.

Let us go back a step and check what would happen if we disabled packing while
performing unroll-and-jam.

\begin{lstlisting}[style=cmd]
// With tiling, unroll-and-jam/unroll, but no packing.
$ mlir-opt -hopt -hopt-copy=false -hopt-unroll -hopt-scalrep -convert-linalg-to-loops
-lower-affine -convert-std-to-llvm dgemm.mlir | mlir-cpu-runner -O3 -e main
-entry-point-result=void -shared-libs=lib/libmlir_runner_utils.so
Compilation time: 0.0568249s \\
11.5595 GFLOPS
\end{lstlisting}

We lose over 2x of the performance if we do not perform packing here! And if
we had not performed the innermost loop unrolling (i.e., set $K_U$ = 1), we get:

\begin{lstlisting}[style=cmd]
// With tiling, packing, and unroll-and-jam/unroll, but K_U = 1.
$ mlir-opt -hopt -hopt-copy -hopt-unroll -hopt-scalrep -convert-linalg-to-loops -lower-affine
-convert-std-to-llvm dgemm.mlir | mlir-cpu-runner -O3 -e main -entry-point-result=void
-shared-libs=lib/libmlir_runner_utils.so
Compilation time: 0.0881901s
20.249 GFLOPS
\end{lstlisting}

All of these improvements are significant from where we started but they are
all finally not enough at all on an absolute scale --- we are still at about
34\% of the peak even after having reproduced a good part of the OpenBLAS/BLIS
recipe!  Clearly, there is something orthogonal that all of this is missing. It
looks like we completely missed vectorization! The highly tuned library kernels
that we have set as a target use a careful selection and schedule of
vector instructions in inline assembly.

Let us see whether our code is even being vectorized reasonably under the hood by
LLVM.  It's pretty straightforward to check this. Instead of looking at the pass
output remarks, we will just look at the generated assembly. The
'-dump-object-file -object-filename' options of mlir-cpu-runner are useful for
this purpose.

\begin{lstlisting}[style=cmd]

$ mlir-opt -hopt -hopt-copy -hopt-unroll -hopt-scalrep -convert-linalg-to-loops
-lower-affine -convert-std-to-llvm dgemm.mlir | mlir-cpu-runner -O3 -e main
-entry-point-result=void -shared-libs=lib/libmlir_runner_utils.so
-dump-object-file -object-filename=hopt.o
$ llvm-objdump -d hopt.o
    ...
    14e5:       c4 e2 a9 b9 e2                   vfmadd231sd     %xmm2, %xmm10, %xmm4
    1856:       c4 42 e5 a8 f8                   vfmadd213pd     %ymm8, %ymm3, %ymm15
    185b:       c4 e3 79 05 a9 30 ff ff ff 01    vpermilpd       $1, -208(%rcx), %xmm5
    1865:       c4 41 7d 28 e6                   vmovapd         %ymm14, %ymm12
    186a:       c4 63 fd 01 f5 55                vpermpd $85, %ymm5, %ymm14
    1870:       c4 62 7d 19 c5                   vbroadcastsd    %xmm5, %ymm8
    1875:       c4 62 e5 a8 c2                   vfmadd213pd     %ymm2, %ymm3, %ymm8
    187a:       c4 42 e5 a8 f5                   vfmadd213pd     %ymm13, %ymm3, %ymm14
    187f:       c5 fd 10 b9 40 ff ff ff          vmovupd -192(%rcx), %ymm7
    1887:       c4 e2 7d 19 b1 40 ff ff ff       vbroadcastsd    -192(%rcx), %ymm6
    1890:       c4 e3 fd 01 d7 ff                vpermpd $255, %ymm7, %ymm2
    1896:       c4 e2 e5 a8 d0                   vfmadd213pd     %ymm0, %ymm3, %ymm2
    189b:       c5 fd 29 93 00 01 00 00          vmovapd %ymm2, 256(%rbx)
    18a3:       c4 e3 fd 01 ef 55                vpermpd $85, %ymm7, %ymm5
    18a9:       c4 63 fd 01 ef aa                vpermpd $170, %ymm7, %ymm13
    18af:       c4 62 e5 a8 ab 80 03 00 00       vfmadd213pd     896(%rbx), %ymm3, %ymm13
    18b8:       c4 e2 e5 a8 e9                   vfmadd213pd     %ymm1, %ymm3, %ymm5
    18bd:       c4 c2 e5 a8 f4                   vfmadd213pd     %ymm12, %ymm3, %ymm6
    18c2:       c5 fb 10 99 60 ff ff ff          vmovsd  -160(%rcx), %xmm3
    18ca:       c5 f9 28 93 20 01 00 00          vmovapd 288(%rbx), %xmm2
    18d2:       c4 62 e9 b9 cb                   vfmadd231sd     %xmm3, %xmm2, %xmm9
    18d7:       c4 c1 7b 10 bc f7 f0 ef ff ff    vmovsd  -4112(%r15,%rsi,8), %xmm7
    18e1:       c4 62 e1 b9 df                   vfmadd231sd     %xmm7, %xmm3, %xmm11
    18e6:       c4 c1 7b 10 84 f7 f0 f7 ff ff    vmovsd  -2064(%r15,%rsi,8), %xmm0
    18f0:       c4 62 e1 b9 d0                   vfmadd231sd     %xmm0, %xmm3, %xmm10
    18f5:       c4 c1 7b 10 4c f7 f0             vmovsd  -16(%r15,%rsi,8), %xmm1
    18fc:       c4 e2 f1 a9 dc                   vfmadd213sd     %xmm4, %xmm1, %xmm3
    1901:       c5 c1 14 e2                      vunpcklpd       %xmm2, %xmm7, %xmm4
    1905:       c5 f1 14 c0                      vunpcklpd       %xmm0, %xmm1, %xmm0
    1909:       c4 e3 7d 18 cc 01                vinsertf128     $1, %xmm4, %ymm0, %ymm1
    190f:       c4 62 7d 19 a1 68 ff ff ff       vbroadcastsd    -152(%rcx), %ymm12
    1918:       c4 42 f5 a8 e7                   vfmadd213pd     %ymm15, %ymm1, %ymm12
    191d:       c4 e3 79 05 a1 70 ff ff ff 01    vpermilpd       $1, -144(%rcx), %xmm4
    1927:       c4 e3 fd 01 c4 55                vpermpd $85, %ymm4, %ymm0
    192d:       c4 e2 7d 19 fc                   vbroadcastsd    %xmm4, %ymm7
    1932:       c4 c2 f5 a8 c6                   vfmadd213pd     %ymm14, %ymm1, %ymm0
    1937:       c4 c2 f5 a8 f8                   vfmadd213pd     %ymm8, %ymm1, %ymm7
    ...
\end{lstlisting}

This is really not the code we would like to see in the innermost loop!  Although
the code is vectorized, it is using XMM registers (128-bit) for a good part instead
of YMM ones (256-bit). More importantly, the FMAs are regularly accompanied by loads
around them and neither should this have been necessary nor is it any good for
performance! A good inner loop here is expected to only have one load of the LHS
and RHS every $N_R$ and $M_R$ FMA ops, roughly speaking, because that's the amount
of reuse we should be getting. In addition, there should be no loads or stores
for the output matrix in the innermost loop, which happens to be the case here.
And finally, it all should have been \%ymm.

It is also important to note that vectorization is often nearly useless unless
the code has been optimized for locality, (i.e., we should be getting data from
caches at least), and it is in fact still not as useful unless we are not reading
from registers most of the time.  We will not dig and analyze any further here to
figure out what went wrong with clang/LLVM and why (this is actually due to a
sub-optimal vectorization choice)!  Instead, we will see how we could get the
vectorization done in MLIR itself where we have all the information and
opportunity.

\subsection{Vectorization}

MLIR supports vector types and vector types as element types for memrefs. Unlike
in LLVM, vector types in MLIR can be multi-dimensional. However, we only need
1-d vector types here. Loop vectorization in MLIR would just lead to IR that
turns memrefs of f64 into memrefs of vector of f64 besides transforming
loops/loop bodies.  A casting op that changes the elemental type of a memref,
and the \href{https://github.com/llvm/llvm-project/tree/master/mlir/docs/Dialects/Standard.md#splat-operation}{splat op}
are also needed here. In more complex cases, the view/subview ops are also
needed. But for this benchmark, the vectorization needed is quite
straightforward, and is a simple outer loop vectorization along the j loop.

The existing ``super vectorizer'' in MLIR is really not functional or complete
for the auto-vectorization we need.  For this article, we build and use a new
loop vectorizer (enabled by -hopt-vect or via -affine-vectorize when running
separately). We introduce a new
\href{https://github.com/bondhugula/llvm-project/commit/6a9822933b28aabced2c6ceef593f35c9665d1fe#diff-862f92fedaa626e6221ac920c18a6b8a}{memref\_shape\_cast op} which is needed to change the elemental type on a memref.
Here's how the vectorized MLIR looks like if we started with the naive matmul
nest.

\begin{lstlisting}[style=mlir]
// mlir-opt -hopt-vect
func @matmul(%A: memref<2048x2048xf64>, %B: memref<2048x2048xf64>, %C: memref<2048x2048xf64>) {
  %0 = memref_shape_cast %B : memref<2048x2048xf64> to memref<2048x512xvector<4xf64>>
  %1 = memref_shape_cast %C : memref<2048x2048xf64> to memref<2048x512xvector<4xf64>>
  affine.for %arg3 = 0 to 2048 {
    affine.for %arg4 = 0 to 512 {
      affine.for %arg5 = 0 to 2048 {
        %2 = affine.load %A[%arg3, %arg5] : memref<2048x2048xf64>
        %3 = splat %2 : vector<4xf64>
        %4 = affine.load %0[%arg5, %arg4] : memref<2048x512xvector<4xf64>>
        %5 = mulf %3, %4 : vector<4xf64>
        %6 = affine.load %1[%arg3, %arg4] : memref<2048x512xvector<4xf64>>
        %7 = addf %5, %6 : vector<4xf64>
        affine.store %7, %1[%arg3, %arg4] : memref<2048x512xvector<4xf64>>
      }
    }
  }
  return
}
\end{lstlisting}

The above IR is generated by vectorizing along the $j$ loop, which is a data
parallel loop as opposed to a reduction dimension). The {\it
memref\_shape\_cast} op casts a memref to one of another shape as long as the
product of the dimension sizes all the way up to any aggregate elemental types
remain the same.  {\it memref\_shape\_cast} actually does not exist in MLIR
trunk.  Also, we do not discuss the case of trip counts not being a multiple of
vector size here since some of the infrastructure around it in MLIR is still
being built. Let us look at the performance with vectorization and with nothing
else enabled.

\begin{lstlisting}[style=cmd]
// Just vectorization.
Compilation time: 0.016561s
1.53043 GFLOPS
\end{lstlisting}

\subsubsection{Rewind and measure with vectorization}

Let us rewind and see how the vectorized code performs step by step, with
tiling, with packing, and with register tiling / unroll-and-jam.

\begin{lstlisting}[style=cmd]
// Just vectorization.
Compilation time: 0.016561s
1.53043 GFLOPS
# Vectorization with tiling
Compilation time: 0.0209861s
7.05307 GFLOPS
\end{lstlisting}

As mentioned earlier, simply vectorizing without worrying about memory bandwidth
gets us nowhere. We have now broken that barrier, and can go further here.

\begin{lstlisting}[style=cmd]
// Vectorization, tiling, and packing/copying
$ mlir-opt -hopt -hopt-vect -hopt-copy -convert-linalg-to-loops -lower-affine
-convert-std-to-llvm dgemm.mlir   | mlir-cpu-runner -O3 -e main -entry-point-result=void -reps=5
-shared-libs=lib/libmlir_runner_utils.so,/usr/local/lib/libblis.so
Compilation time: 0.0409529s
11.2309 GFLOPS
// Vectorization, tiling, packing, and unroll-and-jam, unrolling with MLIR scalar replacement.
$ mlir-opt -hopt -hopt-vect -hopt-copy -hopt-unroll -hopt-scalrep -convert-linalg-to-loops
-lower-affine -convert-std-to-llvm dgemm.mlir | mlir-cpu-runner -O3 -e main -reps=5
-entry-point-result=void -shared-libs=lib/libmlir_runner_utils.so
Compilation time: 0.0383081s
49.8336 GFLOPS
\end{lstlisting}

As observed earlier, once again, the last step has opened the floodgates here
yielding a 4.5x improvement and getting us to 49 GFLOPS -- only 23\% away from
BLIS and about 27\% away from MKL or OpenBLAS.

Let us see how the MLIR looks like right after vectorization, tiling, copying,
and unroll-and-jam (we will disable the $K_C$ loop unroll, i.e., set $K_U$ = 1 to
make it easier to read; the unroll-jamming of $i, j$ is still shown).

\begin{lstlisting}[style=mlir]
affine.for %arg3 = 0 to 8 {
  affine.for %arg4 = 0 to 33 {
    %2 = alloc() : memref<64x256xf64>
    affine.for %arg5 = #map6(%arg4) to min #map7(%arg4) {
      affine.for %arg6 = #map4(%arg3) to #map5(%arg3) {
        %3 = affine.load %arg0[%arg5, %arg6] : memref<2088x2048xf64>
        affine.store %3, %2[%arg4 * -64 + %arg5, %arg3 * -256 + %arg6] : memref<64x256xf64>
      }
    }
    affine.for %arg5 = 0 to 256 {
      %3 = alloc() : memref<256x2xvector<4xf64>>
      affine.for %arg6 = #map4(%arg3) to #map5(%arg3) {
        affine.for %arg7 = #map9(%arg5) to #map10(%arg5) {
          %4 = affine.load %0[%arg6, %arg7] : memref<2048x512xvector<4xf64>>
          affine.store %4, %3[%arg3 * -256 + %arg6, %arg5 * -2 + %arg7] : memref<256x2xvector<4xf64>>
        }
      }
      affine.for %arg6 = #map26(%arg4) to min #map27(%arg4) {
        %4 = alloca() : memref<1xvector<4xf64>>
        %5 = affine.load %1[%arg6 * 4, %arg5 * 2] : memref<2088x512xvector<4xf64>>
        affine.store %5, %4[0] : memref<1xvector<4xf64>>
        %6 = alloca() : memref<1xvector<4xf64>>
        %7 = affine.load %1[%arg6 * 4 + 1, %arg5 * 2] : memref<2088x512xvector<4xf64>>
        affine.store %7, %6[0] : memref<1xvector<4xf64>>
        %8 = alloca() : memref<1xvector<4xf64>>
        %9 = affine.load %1[%arg6 * 4 + 2, %arg5 * 2] : memref<2088x512xvector<4xf64>>
        affine.store %9, %8[0] : memref<1xvector<4xf64>>
        %10 = alloca() : memref<1xvector<4xf64>>
        %11 = affine.load %1[%arg6 * 4 + 3, %arg5 * 2] : memref<2088x512xvector<4xf64>>
        affine.store %11, %10[0] : memref<1xvector<4xf64>>
        %12 = alloca() : memref<1xvector<4xf64>>
        %13 = affine.load %1[%arg6 * 4, %arg5 * 2 + 1] : memref<2088x512xvector<4xf64>>
        affine.store %13, %12[0] : memref<1xvector<4xf64>>
        %14 = alloca() : memref<1xvector<4xf64>>
        %15 = affine.load %1[%arg6 * 4 + 1, %arg5 * 2 + 1] : memref<2088x512xvector<4xf64>>
        affine.store %15, %14[0] : memref<1xvector<4xf64>>
        %16 = alloca() : memref<1xvector<4xf64>>
        %17 = affine.load %1[%arg6 * 4 + 2, %arg5 * 2 + 1] : memref<2088x512xvector<4xf64>>
        affine.store %17, %16[0] : memref<1xvector<4xf64>>
        %18 = alloca() : memref<1xvector<4xf64>>
        %19 = affine.load %1[%arg6 * 4 + 3, %arg5 * 2 + 1] : memref<2088x512xvector<4xf64>>
        affine.store %19, %18[0] : memref<1xvector<4xf64>>
        affine.for %arg7 = 0 to 256 {
          // Unroll-and-jammed body (M_R = 4, N_R = 8 (but vectorized and hence 8/4 effectively).
          %28 = affine.load %2[%arg4 * -64 + %arg6 * 4, %arg7] : memref<64x256xf64>
          %29 = splat %28 : vector<4xf64>
          %30 = affine.load %3[%arg7, 0] : memref<256x2xvector<4xf64>>
          %31 = mulf %29, %30 : vector<4xf64>
          %32 = affine.load %4[0] : memref<1xvector<4xf64>>
          %33 = addf %31, %32 : vector<4xf64>
          affine.store %33, %4[0] : memref<1xvector<4xf64>>
          %34 = affine.load %2[%arg4 * -64 + %arg6 * 4 + 1, %arg7] : memref<64x256xf64>
          %35 = splat %34 : vector<4xf64>
          %36 = mulf %35, %30 : vector<4xf64>
          %37 = affine.load %6[0] : memref<1xvector<4xf64>>
          %38 = addf %36, %37 : vector<4xf64>
          affine.store %38, %6[0] : memref<1xvector<4xf64>>
          %39 = affine.load %2[%arg4 * -64 + %arg6 * 4 + 2, %arg7] : memref<64x256xf64>
          %40 = splat %39 : vector<4xf64>
          %41 = mulf %40, %30 : vector<4xf64>
          %42 = affine.load %8[0] : memref<1xvector<4xf64>>
          %43 = addf %41, %42 : vector<4xf64>
          affine.store %43, %8[0] : memref<1xvector<4xf64>>
          %44 = affine.load %2[%arg4 * -64 + %arg6 * 4 + 3, %arg7] : memref<64x256xf64>
          %45 = splat %44 : vector<4xf64>
          %46 = mulf %45, %30 : vector<4xf64>
          %47 = affine.load %10[0] : memref<1xvector<4xf64>>
          %48 = addf %46, %47 : vector<4xf64>
          affine.store %48, %10[0] : memref<1xvector<4xf64>>
          %49 = affine.load %3[%arg7, 1] : memref<256x2xvector<4xf64>>
          %50 = mulf %29, %49 : vector<4xf64>
          %51 = affine.load %12[0] : memref<1xvector<4xf64>>
          %52 = addf %50, %51 : vector<4xf64>
          affine.store %52, %12[0] : memref<1xvector<4xf64>>
          %53 = mulf %35, %49 : vector<4xf64>
          %54 = affine.load %14[0] : memref<1xvector<4xf64>>
          %55 = addf %53, %54 : vector<4xf64>
          affine.store %55, %14[0] : memref<1xvector<4xf64>>
          %56 = mulf %40, %49 : vector<4xf64>
          %57 = affine.load %16[0] : memref<1xvector<4xf64>>
          %58 = addf %56, %57 : vector<4xf64>
          affine.store %58, %16[0] : memref<1xvector<4xf64>>
          %59 = mulf %45, %49 : vector<4xf64>
          %60 = affine.load %18[0] : memref<1xvector<4xf64>>
          %61 = addf %59, %60 : vector<4xf64>
          affine.store %61, %18[0] : memref<1xvector<4xf64>>
        }
        %20 = affine.load %18[0] : memref<1xvector<4xf64>>
        affine.store %20, %1[%arg6 * 4 + 3, %arg5 * 2 + 1] : memref<2088x512xvector<4xf64>>
        %21 = affine.load %16[0] : memref<1xvector<4xf64>>
        affine.store %21, %1[%arg6 * 4 + 2, %arg5 * 2 + 1] : memref<2088x512xvector<4xf64>>
        %22 = affine.load %14[0] : memref<1xvector<4xf64>>
        affine.store %22, %1[%arg6 * 4 + 1, %arg5 * 2 + 1] : memref<2088x512xvector<4xf64>>
        %23 = affine.load %12[0] : memref<1xvector<4xf64>>
        affine.store %23, %1[%arg6 * 4, %arg5 * 2 + 1] : memref<2088x512xvector<4xf64>>
        %24 = affine.load %10[0] : memref<1xvector<4xf64>>
        affine.store %24, %1[%arg6 * 4 + 3, %arg5 * 2] : memref<2088x512xvector<4xf64>>
        %25 = affine.load %8[0] : memref<1xvector<4xf64>>
        affine.store %25, %1[%arg6 * 4 + 2, %arg5 * 2] : memref<2088x512xvector<4xf64>>
        %26 = affine.load %6[0] : memref<1xvector<4xf64>>
        affine.store %26, %1[%arg6 * 4 + 1, %arg5 * 2] : memref<2088x512xvector<4xf64>>
        %27 = affine.load %4[0] : memref<1xvector<4xf64>>
        affine.store %27, %1[%arg6 * 4, %arg5 * 2] : memref<2088x512xvector<4xf64>>
      }
      dealloc %3 : memref<256x2xvector<4xf64>>
    }
    dealloc %2 : memref<64x256xf64>
  }
}
\end{lstlisting}

The memref<1 x f64> values (single element memrefs) that we see below are a
result of the scalar replacement pass (-affine-scalrep) that runs after the
unroll-and-jam.  We see eight of these summing up to 32 elements of the output
matrix, and these are held in registers when the innermost loop (k) is
iterating. LLVM's passes later turn these into virtual registers.

Both BLIS and OpenBLAS use inline assembly microkernels and other hand written
components that are composed in a modular/reusable way. On the other hand, we
have been trying to generate all our code all the way down, including using
LLVM's instruction
selection, scheduling, and register allocation.  Note that we have got here
through the hop.matmul with the values of $M_C$, $K_C$, $M_R$, $N_R$, $K_U$ indicated
below:

\begin{lstlisting}[style=mlir]
hop.matmul %A, %B, %C {
    M_C = 64 : i32, K_C = 256 : i32, M_R = 4, N_R = 8 : i32, K_U = 4 : i32
} : (memref<2088x2048xf64>, memref<2048x2048xf64>, memref<2088x2048xf64>)
\end{lstlisting}

Before we look at adjusting these for maximum mileage, we will take a detour to
memref layouts.

\subsection{A Quick Detour into Affine Map Layouts}
\label{sec:affine-map-detour}

We are going to take a quick detour here to understand how a memref's layout
map~\cite{memref}
works. A memref type has three pieces of information: its shape, a affine
layout map, and a memory space.

\begin{lstlisting}[style=mlir]
memref<64x256xf64, (d0, d1) -> (d0, d1), 0>
\end{lstlisting}

This is a memref of shape 64x256, and with the identity map (d0, d1) -> (d0,
d1) which corresponds to a row-major layout in memory. 64 rows and 256 columns,
and each of its elements is a float. An element \%A[\%i, \%j] will map to
element
($256*d0 + d1$) in the buffer -- in this case all elements are contiguous. '0'
is the memory space the memref lives in.  On architectures that have different
kinds of explicitly addressable memory, a different memory space id is assigned
to each.  Whenever the layout map is identity, its printing is elided;
similarly, in the case of memory space 0.

Now, let us look at the memref corresponding to the buffer for the LHS matrix
(\%A), which is of type \texttt{memref<64x256xf64>}.

\begin{figure}
  \centering
  \includegraphics[width=0.2\linewidth]{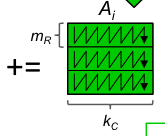}
  \caption{LHS buffer layout.}
\end{figure}

This block is being multiplied with the RHS panel of type
\texttt{memref<256x2xvector<2xf64>>}. The figure above shows how the elements of
A block get traversed: along columns of height $M_R$ all the way up to $K_C$
columns, and then turn around for the next panel, until one has processed all
$M_C$/$M_R$ panels.  Note that the elements are not being accessed contiguously.
But since the block of A is L2 cache resident and is just streamed through L1,
we still get both spatial reuse for it in L1 as well as L2, i.e., it is just
that the reuse in space for A does not happen immediately but once you've
traversed the $M_R$ height.  And we know per this whole approach that it still
stays in cache.  However, there is no harm (and it can only potentially get
better) if we ensure that A is accessed contiguously.  For eg., it can only help
with prefetching into L1.  Hence, we would like to pick a layout different from
the default row-major for \texttt{memref<64x256xf64>}, and MLIR's affine layout
maps make it convenient to express the desired layout here in a compact and
powerful way. If we simply change the memref type to:
\texttt{
memref<64x256xf64, (d0, d1) $\rightarrow$ (d0 floordiv $M_R$, d1, d0 mod $M_R$)>
} \\

it yields the desired layout. The rest of the IR does not need a single
change! The mapping specifies that a $(d0, d1)$ in the logical access space should
be mapped to (d0 flooridv $M_R$, d1, d0 mod $M_R$) in a physical (contiguous) buffer
of size 16x256x4.  When \href{https://github.com/llvm/llvm-project/blob/331c663bd2735699267abcc850897aeaea8433eb/include/mlir/Transforms/Utils.h#L89}{mlir::normalizeMemRef} runs, it will turn this memref
into:

\texttt{
  memref<16x256x4xf64>} \\

And an access Abuf[\%i, \%j] is remapped to Abuf[\%i floordiv 4, \%j, \%i mod 4].
This is an example of a tiled layout, akin to how iteration space tiling is
performed -- stripmine and interchange -- but on the data space.

As another example, here's a memref that accesses a non-contiguous sub-space of
its underlying buffer.
\texttt{
memref<64x256xf64, (d0, d1) -> (d0, d1 floordiv 6, d1 mod 6)>
}
Note that we would have a ``padding'' worth 2 elements (256 \% 6) at the end of each
row that can never be accessed via this memref while staying in bounds. Another
way to write this is:
\texttt{
memref<64x256xf64, $(d0, d1) \rightarrow (d0 * 258 + d1)$>
}

A memref with access strides say 128, 2 (for major, minor respectively) in a larger
underlying buffer can be expressed for example as:
\texttt{
memref<126x100xf32, $(d0, d1) \rightarrow (d0 * 128 + d1 * 2)$>.
}

More examples can be found
\href{https://github.com/llvm/llvm-project/tree/master/mlir/test/Transforms/memref-normalize.mlir}{here}.

The copy options supplied to  mlir::affineDataCopyGenerate  allows one to
choose a custom data layout for the buffer (being copied into/from). One can
choose any layout map as long as it is injective: any injective layout will lead
to semantically correct code.

We will see that this layout is only needed for an
experiment later (Section~\ref{sec:map-to-blis}). We will now
get back to adjusting parameters to go full throttle.

\subsection{Tweaking $M_C$, $K_C$, $M_R$, $N_R$ to maximize reuse}

Using a different setting of these parameters is as simple as changing the
attributes on the hop.matmul op. Note that although one can tune these
parameters on BLIS, its ASM kernel has to be among those that shipped --- these
are long hand-written / hand-unrolled
ones. However, with MLIR we are in a position to also play around with the
register tile sizes just as easily. Let us pick a different set of values of
$M_C$, $K_C$, $M_R$, and $N_R$ for better utilization of registers and L2 cache
capacity as far \%A's buffer goes. We'll set $M_R$ to 3 and $N_R$ to 16 since
this
uses 12 vector registers for \%C instead of the under utilization of 8 earlier
(there are a total of 16 \%ymm registers on x86-64).

\begin{lstlisting}[style=mlir]
hop.matmul %A, %B, %C {
    M_C = 180 : i32, K_C = 480 : i32, M_R = 3, N_R = 16 : i32, K_U = 4 : i32
} : (memref<2088x2048xf64>, memref<2048x2048xf64>, memref<2088x2048xf64>)
\end{lstlisting}

When one chooses $M_R$ = 3 and $N_R$ = 16, it leads to 3 * 16 /4 = 12 256-bit
vector
registers for the output values, 3 values for the LHS, and 1 for the RHS, which
amounts to 12 + 3 + 1 = 16 registers --- exactly the number of VEX encodable
registers for AVX-2. The
BLIS provided kernels for Haswell otherwise
include $6*8$, $8*6$, $4*12$, $12*4$ -- they too use ($2*6$ =) 12 registers for
the output, but these options differ in how much register reuse is relatively
exploited along the two dimensions and how big the L1 resident buffer for the
RHS is. Let us execute the $M_R$ = 3, $N_R$ = 16 configuration.

\begin{lstlisting}[style=cmd]
// M_C = 180 : i32, K_C = 480 : i32, M_R = 3, N_R = 16 : i32, K_U = 4 : i32
$ mlir-opt -hopt -hopt-vect -hopt-copy -hopt-unroll -hopt-scalrep -convert-linalg-to-loops
-lower-affine -convert-std-to-llvm dgemm.mlir | mlir-cpu-runner -O3 -e main -reps=3
-entry-point-result=void -shared-libs=lib/libmlir_runner_utils.so
Compilation time: 0.039474s
61.939 GFLOPS
\end{lstlisting}

We immediately see a 24\% boost in performance! A big
strength of a pure compiler approach here is that one automatically generates
everything -- all the way down to the innermost loop body. So, we have the
opportunity to explore more than what's possible with a fixed set of manually
pre-optimized kernels. This is all under the assumption that the compiler
generated code is competitive or could be made competitive. Let us now explore
the impact of just the $M_R$, $N_R$ values around the best configuration we have
found.

\begin{lstlisting}[style=cmd]
// Let us see the impact of just the M_R, N_R values.
// M_C = 72 : i32, K_C = 256 : i32, M_R = 4, N_R = 8 : i32, K_U = 4 : i32
$ mlir-opt -hopt -hopt-vect  -hopt-copy -hopt-unroll -hopt-scalrep -convert-linalg-to-loops
-lower-affine -convert-std-to-llvm dgemm.mlir | mlir-cpu-runner -O3 -e main -reps=3
-entry-point-result=void -shared-libs=lib/libmlir_runner_utils.so
Compilation time: 0.04356s
49.7282 GFLOPS
// M_R = 6, N_R = 8 (this is what BLIS' micro kernel uses for Haswell).
$ mlir-opt -hopt -hopt-vect -hopt-copy -hopt-unroll -hopt-scalrep -convert-linalg-to-loops
-lower-affine -convert-std-to-llvm dgemm.mlir | mlir-cpu-runner -O3 -e main -reps=3
-entry-point-result=void -shared-libs=lib/libmlir_runner_utils.so
Compilation time: 0.0479391s
40.4135 GFLOPS
// The best conf so far.
// M_C = 180 : i32, K_C = 480 : i32, M_R = 3, N_R = 16 : i32, K_U = 4 : i32.
$ mlir-opt -hopt -hopt-vect -hopt-copy -hopt-unroll -hopt-scalrep -convert-linalg-to-loops
-lower-affine -convert-std-to-llvm dgemm.mlir | mlir-cpu-runner -O3 -e main -reps=3
-entry-point-result=void -shared-libs=lib/libmlir_runner_utils.so
Compilation time: 0.039474s
61.843 GFLOPS
\end{lstlisting}

Let us now peek at the generated assembly of the innermost loop to see how good
it is. A necessary thing is that the innermost block should have no
load/store's on the output (everything should be in registers) and we should
have a continuous sequence of VFMA instructions on 256-bit AVX registers which
are named \%ymm[0-15].

\begin{lstlisting}[style=mlir]
                           # =>        This Inner Loop Header: Depth=5
  vbroadcastsd    -7704(%rbx), %ymm12
  vmovapd -480(%r13), %ymm14
  vfmadd231pd     %ymm12, %ymm14, %ymm0 # ymm0 = (ymm14 * ymm12) + ymm0
  vbroadcastsd    -3864(%rbx), %ymm13
  vfmadd231pd     %ymm13, %ymm14, %ymm1 # ymm1 = (ymm14 * ymm13) + ymm1
  vbroadcastsd    -24(%rbx), %ymm15
  vfmadd213pd     %ymm10, %ymm15, %ymm14 # ymm14 = (ymm15 * ymm14) + ymm10
  vmovapd -448(%r13), %ymm10
  vfmadd231pd     %ymm10, %ymm12, %ymm2 # ymm2 = (ymm12 * ymm10) + ymm2
  vfmadd231pd     %ymm10, %ymm13, %ymm3 # ymm3 = (ymm13 * ymm10) + ymm3
  vfmadd231pd     %ymm10, %ymm15, %ymm4 # ymm4 = (ymm15 * ymm10) + ymm4
  vmovapd -416(%r13), %ymm10
  vfmadd231pd     %ymm10, %ymm12, %ymm5 # ymm5 = (ymm12 * ymm10) + ymm5
  vfmadd231pd     %ymm10, %ymm13, %ymm6 # ymm6 = (ymm13 * ymm10) + ymm6
  vfmadd231pd     %ymm10, %ymm15, %ymm7 # ymm7 = (ymm15 * ymm10) + ymm7
  vmovapd -384(%r13), %ymm10
  vfmadd213pd     %ymm9, %ymm10, %ymm12 # ymm12 = (ymm10 * ymm12) + ymm9
  vfmadd213pd     %ymm11, %ymm10, %ymm13 # ymm13 = (ymm10 * ymm13) + ymm11
  vfmadd231pd     %ymm10, %ymm15, %ymm8 # ymm8 = (ymm15 * ymm10) + ymm8
  vbroadcastsd    -7696(%rbx), %ymm9
  vmovapd -352(%r13), %ymm11
  vfmadd231pd     %ymm9, %ymm11, %ymm0 # ymm0 = (ymm11 * ymm9) + ymm0
  vbroadcastsd    -3856(%rbx), %ymm10
  vfmadd231pd     %ymm10, %ymm11, %ymm1 # ymm1 = (ymm11 * ymm10) + ymm1
  vbroadcastsd    -16(%rbx), %ymm15
  vfmadd213pd     %ymm14, %ymm15, %ymm11 # ymm11 = (ymm15 * ymm11) + ymm14
  vmovapd -320(%r13), %ymm14
  vfmadd231pd     %ymm14, %ymm9, %ymm2 # ymm2 = (ymm9 * ymm14) + ymm2
  vfmadd231pd     %ymm14, %ymm10, %ymm3 # ymm3 = (ymm10 * ymm14) + ymm3
  vfmadd231pd     %ymm14, %ymm15, %ymm4 # ymm4 = (ymm15 * ymm14) + ymm4
  vmovapd -288(%r13), %ymm14
  vfmadd231pd     %ymm14, %ymm9, %ymm5 # ymm5 = (ymm9 * ymm14) + ymm5
  vfmadd231pd     %ymm14, %ymm10, %ymm6 # ymm6 = (ymm10 * ymm14) + ymm6
  vfmadd231pd     %ymm14, %ymm15, %ymm7 # ymm7 = (ymm15 * ymm14) + ymm7
  vmovapd -256(%r13), %ymm14
  vfmadd213pd     %ymm12, %ymm14, %ymm9 # ymm9 = (ymm14 * ymm9) + ymm12
  vfmadd213pd     %ymm13, %ymm14, %ymm10 # ymm10 = (ymm14 * ymm10) + ymm13
  vfmadd231pd     %ymm14, %ymm15, %ymm8 # ymm8 = (ymm15 * ymm14) + ymm8
  vbroadcastsd  -7688(%rbx), %ymm12
  vmovapd  -224(%r13), %ymm14
  ...
\end{lstlisting}

This looks as good as we may expect! The output matrix values are always in
registers (no spilling). LHS and RHS values have the intended amount of reuse in
registers. The LHS is loaded in once via broadcast/splat
(vbroadcastsd), and reused along the $j$ register tile dimension, while the RHS is
moved in via aligned vector loads and reused along the $i$ register tile
dimension.

\begin{lstlisting}[style=cmd]
// Let us look at the benefit of packing in isolation on the best code we have.
$ mlir-opt -hopt -hopt-vect -hopt-copy -hopt-unroll -hopt-scalrep -convert-linalg-to-loops
-lower-affine -convert-std-to-llvm dgemm.mlir | mlir-cpu-runner -O3 -e main -reps=3
-entry-point-result=void -shared-libs=lib/libmlir_runner_utils.so
Compilation time: 0.039474s
61.843 GFLOPS
\end{lstlisting}
\begin{lstlisting}[style=cmd]
// Without packing (but with everything else).
$ mlir-opt -hopt -hopt-vect -hopt-copy=false -hopt-unroll -hopt-scalrep -convert-linalg-to-loops
-lower-affine -convert-std-to-llvm dgemm.mlir | mlir-cpu-runner -O3 -e main
-entry-point-result=void -shared-libs=lib/libmlir_runner_utils.so,/usr/local/lib/libblis.so
Compilation time: 0.030535s
22.5257 GFLOPS
// A 3x drop in performance just due to the lack of packing, while it was just
// 1.5x on a code that was not fully optimized.
\end{lstlisting}
\begin{lstlisting}[style=cmd]
// Without unroll-and-jam (but with packing and everything else)
$ mlir-opt -hopt -hopt-vect -hopt-copy -hopt-unroll=false -hopt-scalrep -linalg-lower-to-loops
-linalg-convert-to-llvm -convert-linalg-to-loops -lower-affine -convert-std-to-llvm
dgemm.mlir | mlir-cpu-runner -O3 -reps=5 -e main -entry-point-result=void
-shared-libs=lib/libmlir_runner_utils.so,/usr/local/lib/libblis.so
Compilation time: 0.0424139s
15.5651 GFLOPS
\end{lstlisting}
\begin{lstlisting}[style=cmd]
// Without packing and without unroll-and-jam (but with everything else)
$ mlir-opt -hopt -hopt-vect -hopt-copy=false -hopt-unroll=false -hopt-scalrep
-linalg-lower-to-loops -linalg-convert-to-llvm -convert-linalg-to-loops -lower-affine
-convert-std-to-llvm dgemm.mlir | mlir-cpu-runner -O3 -reps=5 -e main
-entry-point-result=void
-shared-libs=lib/libmlir_runner_utils.so,/usr/local/lib/libblis.so
Compilation time: 0.0228369s
10.785 GFLOPS
\end{lstlisting}

As for $M_C$, $K_C$, note that we are using large enough values that the $M_C$ x $K_C$
tile (675 KB for $M_C$ = 180, $K_C$ = 480) of the LHS matrix (A) actually fits in
the L3 cache, as opposed to in the L2 cache, which was the plan with the
BLIS/OpenBLAS tiling approach.  Since we haven't really done the additional
level of tiling ($N_C)$ per the OpenBLAS/BLIS strategy, we notice that we get more
mileage here out of using a larger tile for the LHS in L3 instead of a smaller
tile in L2 (the L3 was sort of unoccupied for us). Note that a larger $M_C$ in
particular amortizes the cost of the RHS's transfer into L1 and exploits more L1
reuse on that tile.  The additional level of tiling is conceptually
straightforward given where we are now, and this could be evaluated along with
better values for $M_C$, $K_C$ for improved L2 and L3 reuse. We skip that for
this article.

Let us now look back at our journey up until this best performing result, and the
role each optimization played.
\begin{figure}[!h]
    \centering
    \def\svgwidth{\columnwidth}
    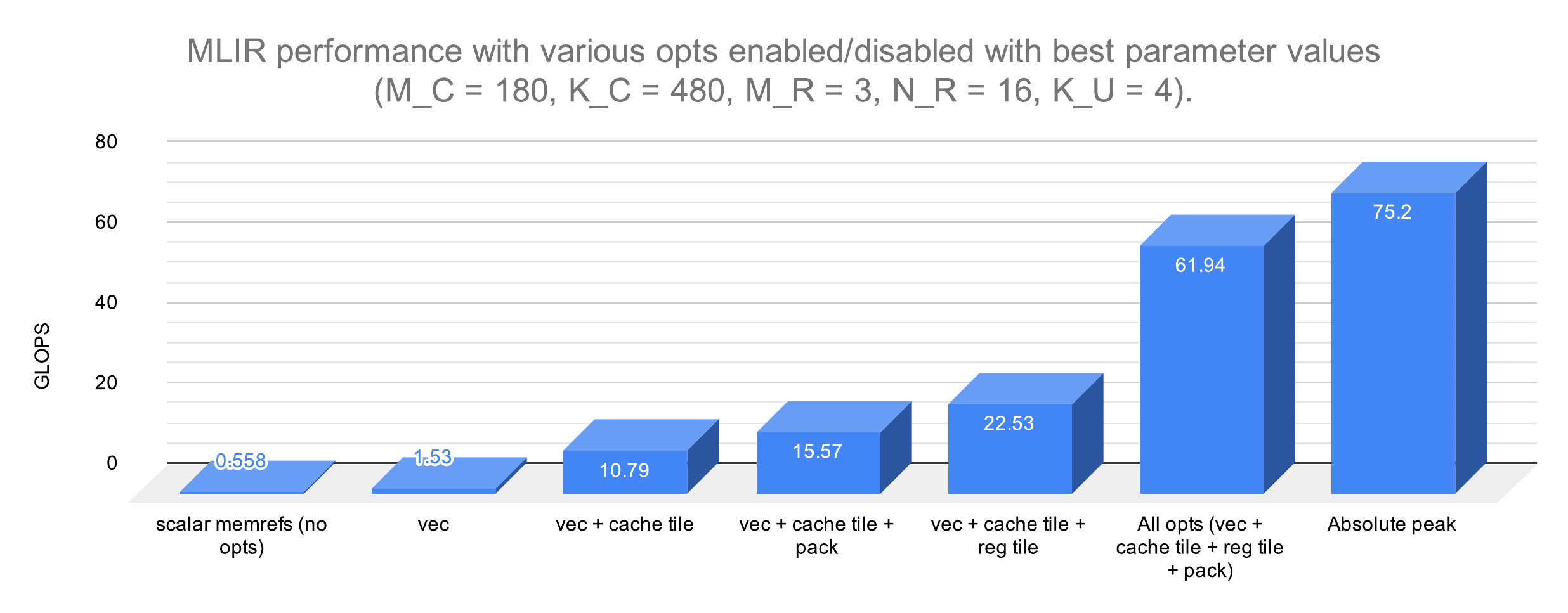
\end{figure}

\subsection{Within 9\% of OpenBLAS and MKL}

Let us plot the overall picture so far.
\begin{figure}[!h]
    \centering
    \def\svgwidth{\columnwidth}
    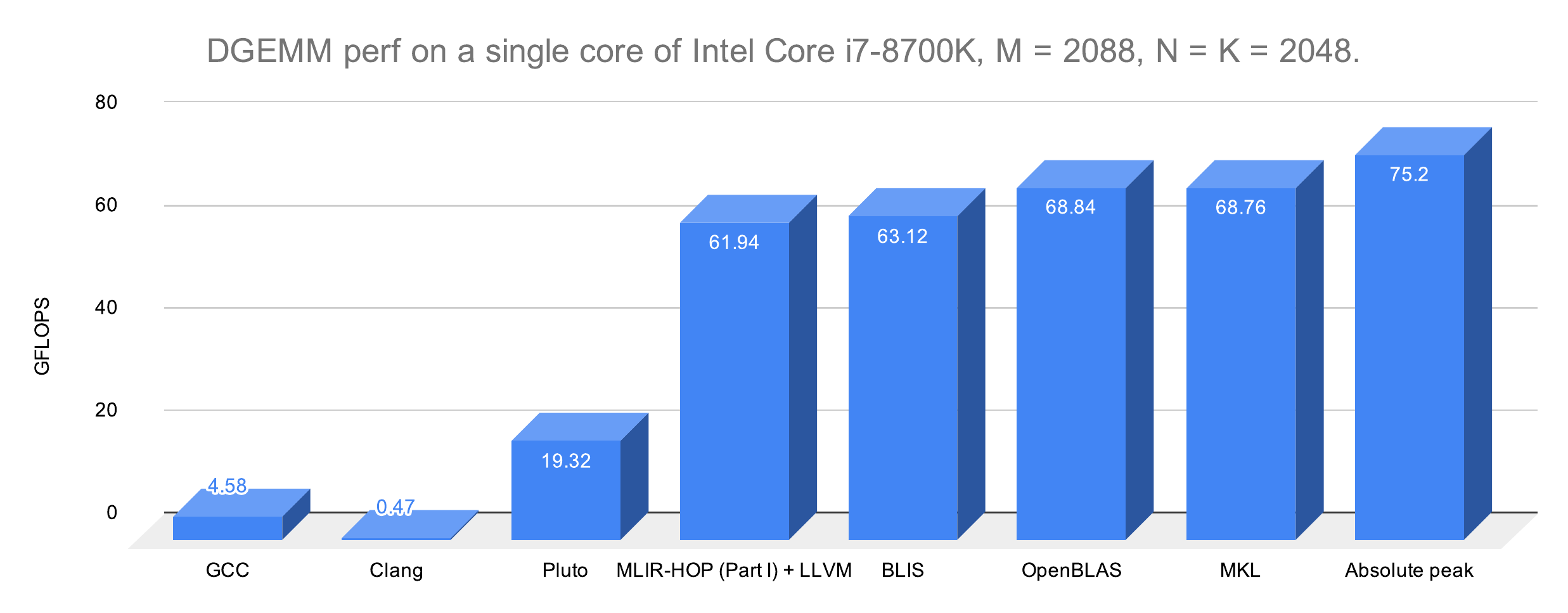
\end{figure}

We have now nearly matched BLIS performance (61 vs 63 GFLOPS), and are only 9%
away from MKL and OpenBLAS's performance! Note that BLIS' performance could
also likely improved by tuning and playing around $M_C$ and $K_C$ as opposed to
choosing the configuration it preset ($M_C$ = 72, $K_C$ = 256), but as we
observed here, it is the $M_R$, $N_R$ values that have a greater impact here as
far as the MLIR + LLVM approach goes. Secondly, further tuning of BLIS is likely
to still be around the OpenBLAS performance (based on published results).

\subsubsection{Vectorization with copying, unroll-and-jam, unrolling but no
scalar replacement}

On this note, let us look at what happens if we disabled MLIR's scalar
replacement but enabled unroll-and-jam, sort of leaving it to LLVM to perform
the replacement.

\begin{lstlisting}[style=cmd]
$ mlir-opt -hopt -hopt-vect -hopt-scalrep=false -convert-linalg-to-loops -lower-affine
-convert-std-to-llvm dgemm.mlir | mlir-cpu-runner -O3 -e main -entry-point-result=void -reps=5
-shared-libs=lib/libmlir_runner_utils.so,/usr/local/lib/libblis.so
Compilation time: 0.038455s
10.7969 GFLOPS
\end{lstlisting}
Relying on LLVM to do the scalar replacement does not work! We will not again go
into why. Although the passes in LLVM could be improved here, this is an
optimization that MLIR is supposed to perform well since it has to do with
multidimensional subscripts and loops.

\subsection{Map to BLIS micro-kernel: ``Outer MLIR, Inner BLIS''}
\label{sec:map-to-blis}
We might wonder at this point as to what would happen if the MLIR approach just
borrowed the BLIS micro-kernel for its inner three loops (the ones running with
trip counts $K_C$, $M_R$, $N_R$). How much better performance would it lead to for the
best MLIR version?  Given that we were able to get close to BLIS using MLIR +
LLVM all the way, a no difference in performance would be great news
for those who have worked on the x86/x86-64 LLVM backends and most of the low
level infrastructure surrounding it.

We will do a simple experiment here. One could imagine employing BLIS'
micro-kernel (multiplication of two panels: $M_R$ x $K_C$  and $K_C$ x $N_R$) within
MLIR itself, i.e., the outer loops, tiling, explicit copying
etc. are all automated with MLIR the way we described above, but the carefully
crafted inline assembly kernel of BLIS is called for the inner part. This is
also an approach of high interest to several research groups working on HPC
compilers today because many believe that (1) the performance of the
hand-written inner kernel cannot be attained through compiler generated code, (2)
one shouldn't be subject to the vagaries of a compiler when generating such code
where even the last ounce of performance should be extracted. Both points are
debatable, and the issues involved are addressable.

Over here, our experiment is also interesting because it tells us whether the
remaining difference in performance is coming from the carefully tailored
instruction selection and schedule used in BLIS' micro-kernel, which the
compiler (having been tasked to deal with the entire world of programs) is
unable to nail down optimally. Seeing a big difference when we swap out our
inner loops with the BLIS kernel could be disappointing because it would mean
we will have to now get into LLVM to see how we could improve things as one
option. On the other hand, the lack of any significant improvement would be
great news for us because we could then focus on the "macro kernel", code and
loops that we have all possible control over, to go closer to the peak. It
would also be a tribute to all those who have
worked on LLVM backends and surrounding infrastructure.

\begin{figure}[!h]
  \centering
  \includegraphics[width=0.8\linewidth]{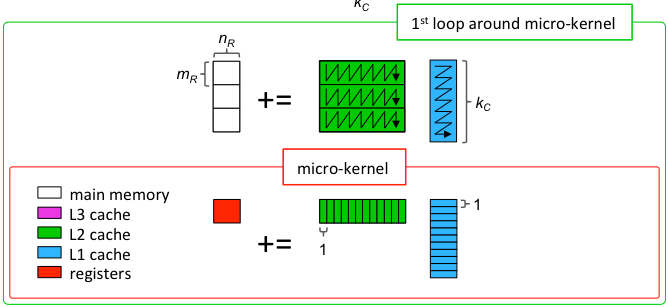}
  \caption{The BLIS micro-kernel (figure courtesy: Field Van
  Zee and Robert van de Geijn). }
\end{figure}

The BLIS micro-kernel being used here is
\href{https://github.com/flame/blis/blob/b426f9e04e5499c6f9c752e49c33800bfaadda4c/kernels/haswell/3/bli_gemm_haswell_asm_d6x8.c#L926}{
bli\_dgemm\_haswell\_asm\_6x8}; this
corresponds to $M_R$ = 6 and $N_R$ = 8. We'll thus run a pure MLIR code
generated
version with $M_C$ = 174, $K_C$ = 512, $M_R$ = 6, $N_R$ = 8, and then the same
one with
the innermost 3-d nest mapped to BLIS' microkernel. This keeps the comparison
meaningful. (We have adjusted 180 and 480 slightly to 174 and 512 respectively to
ensure that all cache tiles are full tiles to keep things more readable).

We also need to change the layout of the LHS buffer to the packed layout shown
in green - this was discussed in detail earlier when we described how memref
layouts work (Section~\ref{sec:affine-map-detour}).

Let us look at the structure of MLIR here with tiling and packing, but without
vectorization or any unroll-jamming enabled since the inner portion is going to
be swapped out.

\begin{lstlisting}[style=mlir]
affine.for %arg1 = 0 to 4 {
  affine.for %arg2 = 0 to 12 {
    %9 = alloc() : memref<29x512x6xf64>
    affine.for %arg3 = #map4(%arg2) to #map5(%arg2) {
      affine.for %arg4 = #map2(%arg1) to #map3(%arg1) {
        %10 = affine.load %0[%arg3, %arg4] : memref<2088x2048xf64>
        affine.store %10, %9[%arg2 * -29 + %arg3 floordiv 6, %arg1 * -512 + %arg4, %arg3 mod 6]
											: memref<29x512x6xf64>
      }
    }
    affine.for %arg3 = 0 to 256 {
      %10 = alloc() : memref<512x8xf64>
      affine.for %arg4 = #map2(%arg1) to #map3(%arg1) {
        affine.for %arg5 = #map7(%arg3) to #map8(%arg3) {
          %11 = affine.load %1[%arg4, %arg5] : memref<2048x2048xf64>
          affine.store %11, %10[%arg1 * -512 + %arg4, %arg3 * -8 + %arg5] : memref<512x8xf64>
        }
      }
      affine.for %arg4 = #map15(%arg2) to #map16(%arg2) {
        affine.for %arg5 = 0 to 512 {
          affine.for %arg6 = 0 to 8 {
            %11 = affine.load %10[%arg5, %arg6] : memref<512x8xf64>
            affine.for %arg7 = 0 to 6 {
              %12 = affine.load %9[%arg2 * -29 + %arg4 + %arg7 floordiv 6, %arg5, %arg7 mod 6]
									 : memref<29x512x6xf64>
              %13 = affine.load %2[%arg4 * 6 + %arg7, %arg3 * 8 + %arg6] : memref<2088x2048xf64>
              %14 = mulf %12, %11 : f64
              %15 = addf %14, %13 : f64
              affine.store %15, %2[%arg4 * 6 + %arg7, %arg3 * 8 + %arg6] : memref<2088x2048xf64>
            }
          }
        }
      }
      dealloc %10 : memref<512x8xf64>
    }
    dealloc %9 : memref<29x512x6xf64>
  }
}
\end{lstlisting}

Note that the buffer of shape <174x512xf64> after being assigned the tiled
layout and post memref normalization has been converted to the shape
<29x512x6xf64>, and it has elements exactly in the order the BLIS micro-kernel
wants. One can notice that
the blocks and panels corresponding to the buffers here being multiplied (LHS
buffer: <29x512x6xf64>, RHS buffer: <512x8xf64>, the output is just
2048x2048xf64 since we will reuse it in registers). For the purpose of this
article, we developed a -hopt-blis pass option that actually maps the right part
to the BLIS micro-kernel.  For a 2088x2048 matrix, this is the generated IR:

\begin{lstlisting}[style=mlir]
affine.for %arg1 = 0 to 4 {
  affine.for %arg2 = 0 to 12 {
    %9 = alloc() {alignment = 32 : i32} : memref<29x512x6xf64>
    affine.for %arg3 = #map4(%arg2) to #map5(%arg2) {
      affine.for %arg4 = #map2(%arg1) to #map3(%arg1) {
        %10 = affine.load %0[%arg3, %arg4] : memref<2088x2048xf64>
        affine.store %10, %9[%arg2 * -29 + %arg3 floordiv 6, %arg1 * -512 + %arg4, %arg3 mod 6]
                      : memref<29x512x6xf64>
      }
    }
    affine.for %arg3 = 0 to 256 {
      %10 = alloc() {alignment = 32 : i32} : memref<512x8xf64>
      affine.for %arg4 = #map2(%arg1) to #map3(%arg1) {
        affine.for %arg5 = #map7(%arg3) to #map8(%arg3) {
          %11 = affine.load %1[%arg4, %arg5] : memref<2048x2048xf64>
          affine.store %11, %10[%arg1 * -512 + %arg4, %arg3 * -8 + %arg5] : memref<512x8xf64>
        }
      }
      call @hopt_dgemm_blis_kernel(%9, %10, %2, %c512, %arg2, %arg3)
          : (memref<29x512x6xf64>, memref<512x8xf64>, memref<2088x2048xf64>, index, index, index) -> ()
      dealloc %10 : memref<512x8xf64>
    }
    dealloc %9 : memref<29x512x6xf64>
  }
}
\end{lstlisting}

The hopt\_dgemm\_blis\_kernel function is added to
mlir\_runtime\_utils, and just wraps around
\href{https://github.com/flame/blis/blob/b426f9e04e5499c6f9c752e49c33800bfaadda4c/kernels/haswell/3/bli_gemm_haswell_asm_d6x8.c#L926}{bli\_dgemm\_haswell\_asm\_6x8}.

The allocs above have alignments since the BLIS kernel uses 256-bit aligned
loads on these buffers. So, this experimentation was done with additional
support in the MLIR's std to llvm dialect conversion to use aligned\_alloc.
We have these alignments set even for the pure MLIR generated code since the
loads and stores on vector types will have the vector size as the default ABI
alignment, leading to aligned load/stores during LLVM code
generation; so the alloc's need to be aligned to vector size boundaries.

\begin{lstlisting}[style=cmd]
// MLIR with BLIS micro-kernel: M_R = 6, N_R = 8.
$ mlir-opt -hopt -hopt-blis -convert-linalg-to-loops -lower-affine -convert-std-to-llvm dgemm.mlir
| mlir-cpu-runner -O3 -e main -entry-point-result=void
-shared-libs=lib/libmlir_runner_utils.so,/usr/local/lib/libblis.so -dump-object-file
-object-filename=hopt.o
Compilation time: 0.0281591s
61.421 GFLOPS
\end{lstlisting}
\begin{lstlisting}[style=cmd]
// MLIR with pure codegen M_R = 6, N_R = 8.
$ mlir-opt -hopt -convert-linalg-to-loops -lower-affine -convert-std-to-llvm dgemm.mlir |
mlir-cpu-runner  -O3 -e main -reps=5  -entry-point-result=void
-shared-libs=lib/libmlir_runner_utils.so
Compilation time: 0.0475061s
40.2426 GFLOPS
\end{lstlisting}
\begin{lstlisting}[style=cmd]
// MLIR with pure codegen M_R = 4, N_R = 8.
$ mlir-opt -hopt  -convert-linalg-to-loops -lower-affine -convert-std-to-llvm dgemm.mlir |
mlir-cpu-runner  -O3 -e main -reps=5  -entry-point-result=void
-shared-libs=lib/libmlir_runner_utils.so
Compilation time: 0.0415299s
50.7075 GFLOPS
\end{lstlisting}
\begin{lstlisting}[style=cmd]
// Recall with M_R = 3, N_R = 16.
$ mlir-opt -hopt -hopt-copy -hopt-unroll -hopt-scalrep -convert-linalg-to-loops -lower-affine
-convert-std-to-llvm dgemm.mlir | mlir-cpu-runner -O3 -e main -reps=3 -entry-point-result=void
-shared-libs=lib/libmlir_runner_utils.so
Compilation time: 0.039474s
61.939 GFLOPS
\end{lstlisting}

There is nearly no difference when comparing the hybrid MLIR and BLIS micro
kernel with the best MLIR version we had. For a given $M_R$ x $N_R$ register tile
with schedule we have been using, the register requirement (assuming the
instruction scheduler is not doing any non-trivial reordering) would be
$M_R*N_R/4 + M_R$ + 1 (the division by four is for the vector width). With $M_R =
3$, $N_R = 16$, this requirement would exactly be 16, which is the number of \%ymm
registers we have! Using $M_R$ = 6, $N_R$ = 8 with our code (as opposed to with the
BLIS micro-kernel), the requirement goes up to 6\*2 + 6 + 1 = 19
registers, and leads to a spill. The assembly dump confirms this
(notice the vmovupd stores in between the FMAs, which didn't exist earlier).
Also, the intuitively sub-optimal sizes of $M_R$ = 4, $N_R$ = 8 provide better
performance than the tighter $M_R$ = 6, $M_R$ = 8, indicating the cliff associated
with spilling.

\begin{lstlisting}[style=mlir]
c4 62 95 b8 f0                                vfmadd231pd     %ymm0, %ymm13, %ymm14
1360:       c4 a2 7d 19 84 ef e8 c3 ff ff     vbroadcastsd    -15384(%rdi,%r13,8), %ymm0
136a:       c5 fd 11 84 24 80 01 00 00        vmovupd         %ymm0, 384(%rsp)
1373:       c4 e2 95 b8 c8                    vfmadd231pd     %ymm0, %ymm13, %ymm1
1378:       c4 a2 7d 19 84 ef e8 d2 ff ff     vbroadcastsd    -11544(%rdi,%r13,8), %ymm0
1382:       c5 fd 11 84 24 a0 01 00 00        vmovupd %ymm0, 416(%rsp)
138b:       c4 e2 95 b8 d0                    vfmadd231pd     %ymm0, %ymm13, %ymm2
1390:       c4 a2 7d 19 84 ef e8 e1 ff ff     vbroadcastsd    -7704(%rdi,%r13,8), %ymm0
139a:       c4 e2 95 b8 d8                    vfmadd231pd     %ymm0, %ymm13, %ymm3
139f:       c4 22 7d 19 a4 ef e8 f0 ff ff     vbroadcastsd    -3864(%rdi,%r13,8), %ymm12
13a9:       c5 7d 11 a4 24 e0 01 00 00        vmovupd %ymm12, 480(%rsp) 13b2:
            c4 c2 95 b8 e4                    vfmadd231pd     %ymm12, %ymm13, %ymm4
\end{lstlisting}

However, $M_R$ = 6, $N_R$ = 8 could be made to fit in the register budget by using a
different permutation of instructions for the innermost loop body, i.e., with
a register tile whose intra-tile loops have been permuted before the unrolling
was performed. In such a case, the register requirement would be: $M_R$ * $N_R$/4 +
$N_R$/4 + 1 = 15! The schedule to be used to realize that would be:

\[
(d0, d1, d2) \rightarrow (d2 \textrm{ floordiv } 480, d0 \textrm{ floordiv } 330, d1 \textrm{ floordiv } 8, d0 \textrm{ floordiv } 6,
d2, d0, d1).
\]

Notice the flip $(\dots, d0, d1)$ instead of $(\dots, d1, d0)$. This also means that the
LLVM backend is not going to automatically do a complex permutation on the
innermost body that changes live ranges, reducing register requirements to
fit within the budget and thereby avoiding a spill. Using the above schedule
leads to no spill and the code performs at about 5\% slower though than $M_R$ = 3,
$N_R$ = 16. It is thus also important to get the right permutation for the
unrolled register tile (at the time of selecting the loop transformation) --- so
that the matrix values with longer reuse distances for that intra register tile
permutation (LHS values in the case of (d1, d0) and RHS values in the case of
(d0, d1)) do not cause a spill. Alternatively, this problem could be viewed as
one that performs the unroll-and-jam for $i$, $j$ in the right order. The register
requirement in the context of this
"known" computation and its interaction with the intra-tile loop permutation and
register tile sizes ($M_R$, $N_R$) are pretty easy to capture at a high level to
make sure even a simple register allocator does the right thing.

Overall, we conclude here that the kind of code we could get automatically
is clearly on par or close to what was achieved with expert written assembly.
The best performing $M_R$x$N_R$ of 3x16 is on par with the MLIR-BLIS configuration
we explored. The 6x8 version, if made to work without spilling, will likely be
on par with 3x16.  This basically means the 4-9\% difference between the pure
MLIR one and the BLIS/OpenBLAS ones likely stems from things around and outside
the micro-kernel. This part requires more experimentation before any firm
conclusions can be made.

On a separate note, one can see that MLIR does make it easier to map to external
hand-optimized kernels, and combine the best of both worlds if that's necessary
for peak performance. Also, it is straightforward to perform a systematic
exploration around interesting values by generating ops with different
parameter attributes.  However, each choice requires a compile and execute
cycle since we are not generating any target code parametric in the identified
parameters.

\begin{figure}[!h]
	\centering
  \def\svgwidth{\columnwidth}
	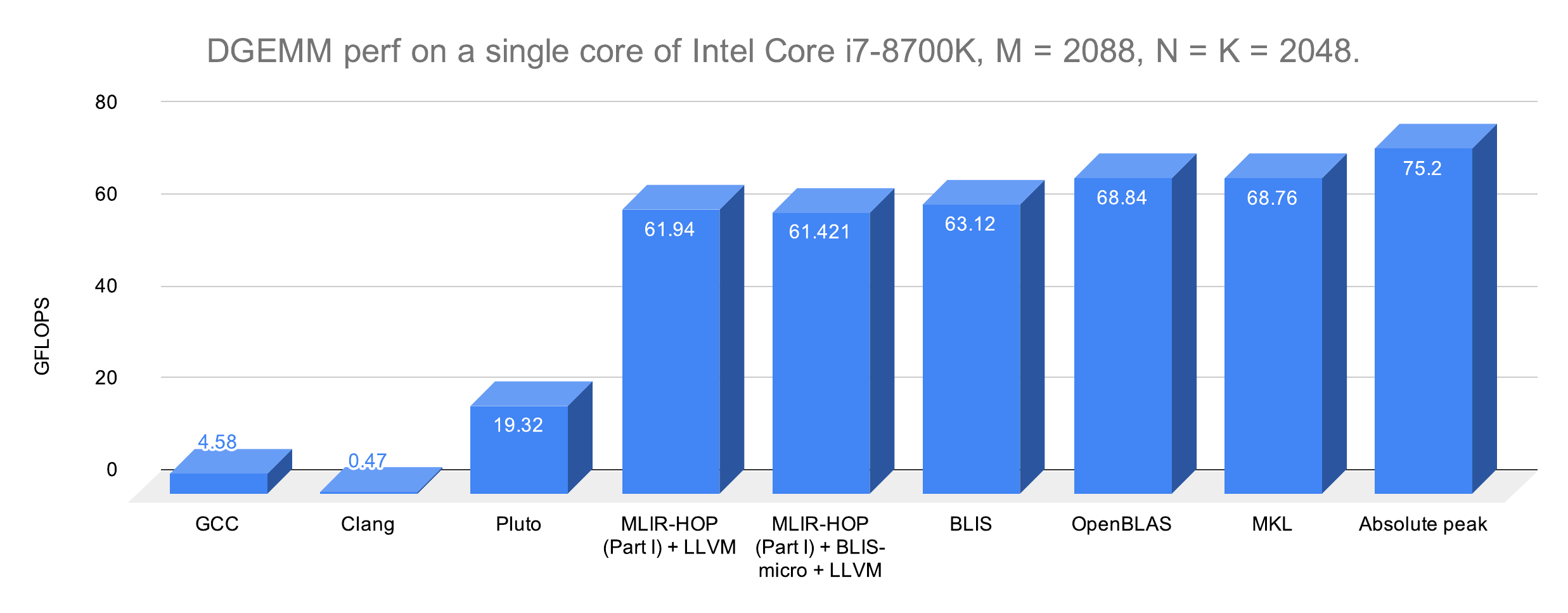
	\caption{DGEMM performance.}
\end{figure}

\FloatBarrier

\subsection{SGEMM performance}

We can similarly now benchmark SGEMM performance quickly. All that we need to
change on the op's operands is an s/f64/f32! In addition, we will just double
the register tile size $N_R$ from 16 to 32 since two times the number of f32
elements can be held in a vector register (8 x f32 instead of 4 x f64). We would
thus still be using 3x4 = 12 vector registers for C. In addition, we will also
double $M_C$ to 348 for the same reason. We thus use this op to generate SGEMM.

\begin{lstlisting}[style=mlir]
hop.matmul %A, %B, %C {
    M_C = 348 : i32, K_C = 512 : i32, M_R = 3, N_R = 32 : i32, K_U = 4 : i32
} : (memref<2088x2048xf32>, memref<2048x2048xf32>, memref<2088x2048xf32>)
\end{lstlisting}

Evaluating this, we find that the performance of purely MLIR generated code is
within 2\% of MKL performance here!, and in fact marginally better than OpenBLAS
and BLIS. The outer MLIR + inner BLIS version here delivers the expected
performance, nearly on par with pure MLIR here.

\FloatBarrier

\begin{figure}[!ht]
	\centering
	\def\svgwidth{\columnwidth}
	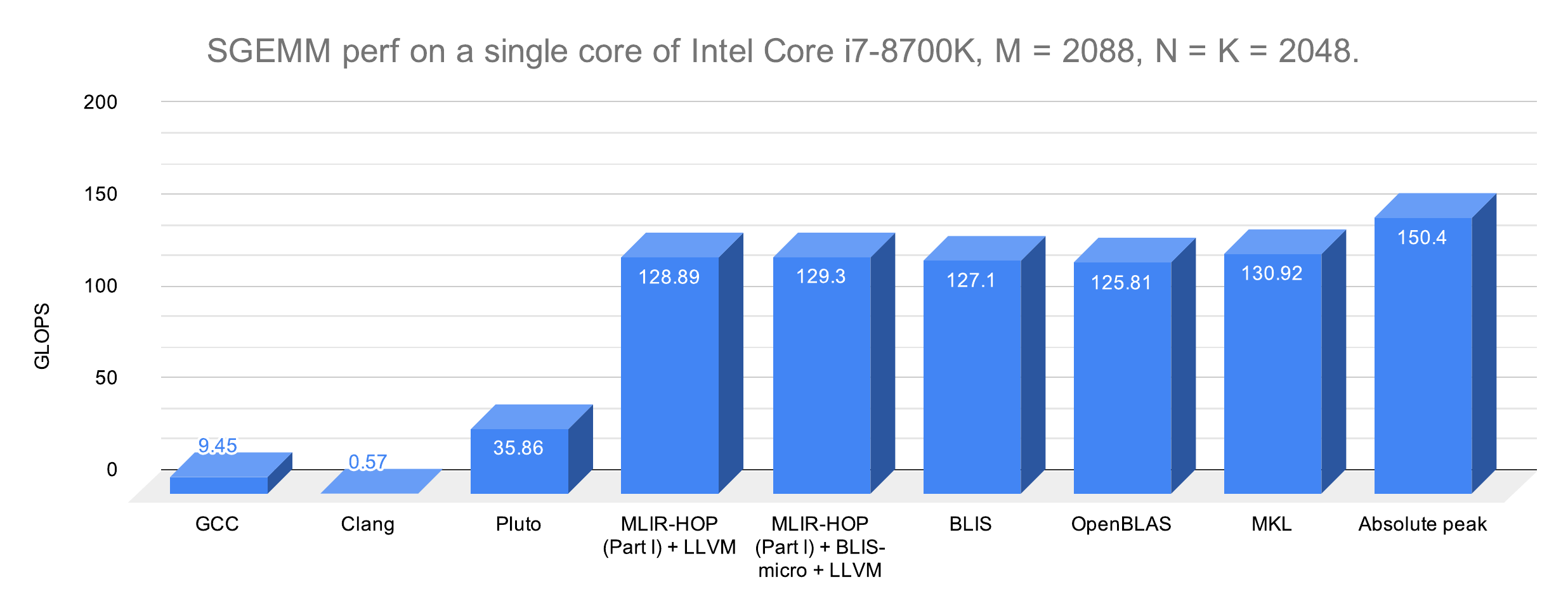
  \caption{SGEMM performance.}
\end{figure}

\subsection{What about the remaining 9\%?}

At this point, we would like to see what stands in the way of bridging the
remaining 9\% gap in getting to MKL/OpenBLAS's performance. In particular, the
hand-written kernels in OpenBLAS, BLIS also use prefetching -- we are not
yet generating any of these instructions.

In summary, we have been able to generate all code starting from just this op:
\begin{lstlisting}[style=mlir]
hop.matmul %A, %B, %C {
    M_C = 180 : i32, K_C = 480 : i32, M_R = 3, N_R = 16 : i32, K_U = 4 : i32
} : (memref<2088x2048xf64>, memref<2048x2048xf64>, memref<2088x2048xf64>)
\end{lstlisting}

\subsection{Other questions}

There may be a few questions here.

\begin{enumerate}

\item While the code was automatically generated, the transformation sequence was
specified from within an MLIR pass. What does the sequence look like - just C++
IR building and calls to utilities/passes? How productive is it and the
subsequent exploration?

\item What about problem sizes that are not a perfect multiple of register tile
sizes ($M_R$, $N_R$)? While MLIR's unroll-and-jam works with cleanup code
generated, there are unresolved issues in interactions downstream (for
performance). There is literature in the polyhedral body of works on separation
of full and partial tiles, and this is largely an implementation issue. Note
that cache tile sizes need not perfectly divide problem sizes - this already
works in MLIR in conjunction with packing and other transformations presented
here. For eg. the best $M_C$ size we used (180) does not divide 2048.

\item We have primarily used the polyhedral passes in MLIR here since the generated
code at each stage always stays affine.  The
    \href{https://github.com/llvm/llvm-project/tree/master/mlir/test/Transforms/memref-normalize.mlir}{LinAlg
    dialect} does not have the utilities to automatically analyze and generate
    packing code
for example.  Nearly all passes and utilities used here such as unroll-and-jam,
  scalar replacement, and memref normalization work on affine dialect ops.

\item What about all the other options around input matrices such as strides and
transpose? These all are cleanly expressed via memrefs' affine layout map
discussed above (Section~\ref{sec:affine-map-detour}). We haven't considered the
    $\alpha$ and $\beta$ arguments for DGEMM though (it is really $C =
    \alpha\*A\*B + \beta\*C$) - we have actually assumed $\alpha$ = $\beta$ = 1.
    In fact, MLIR's pattern rewrites can fold and optimize away code/nests when
    $\alpha$ or $\beta$ are 0 or 1.  Overall, these aspects are something to
    keep in mind before making a complete/exact comparison.

\item How do we determine good or the optimal parameter values? The work of
    \href{https://dl.acm.org/citation.cfm?id=2925987}{Low et al.} is on
analytical modeling to derive good parameter values (the derived formulae are
pretty straightforward to just plug values from hardware resources into).
Besides such analytical modeling, there is a lot of room to play here depending
on how powerful the generative infrastructure is, and to what extent we would
like to generalize this approach beyond the domain considered here.

\end{enumerate}

\section{Related Work}

Most of the related work on OpenBLAS~\cite{goto2008toms} and
BLIS~\cite{vanzee2015toms} was already described inline. The work by
Gareev et al.~\cite{gareev18taco}  is the only approach I am aware of that
tried the BLIS approach from inside a compiler (Polly/LLVM there). From their
results, they appeared to have had reached about 70\% of MKL/OpenBLAS
performance.  With a low-level IR such as LLVM, and with Polly and LLVM
infrastructure decoupled to some extent, one's expressive power and mileage
is expected to vary. There has
been a large amount of work on empirical search driven library generation for
GEMM~\cite{atlas} (ATLAS), and works like that of Yotov et
al.~\cite{yotov2005ieee} have plugged in analytical models to drive such
optimization with comparable performance. The work of Low et
al.~\cite{low2016toms} makes a similar attempt with BLIS. However, all of these
frameworks were centered around generation of C code along with inner inline
assembly kernels.

\section{Reproducing these Results}

\paragraph{Software setup:} Fedora Linux 30 running kernel
5.3.6-200.fc30.x86\_64, BLIS version 0.6.0-40-gf4f5170f, MKL version 2019.4.243,
OpenBLAS 0.3.7-1.fc30.x86\_64, and Pluto git 0.11.4-903-g7f21ab57. {\it
cpupower} was used to set the frequency governor to `performance'.

A good part of the experiments presented in this article can be reproduced with
MLIR trunk. There are major features though that are pending upstream
integration (memref\_shape\_cast op, alloca op, scalar replacement, a new
vectorization pass/utility, and support for a few packing options), but these
are available in the \href{
  https://github.com/bondhugula/llvm-project/tree/hop}{$hop$ branch here}, which
is the
recommended way of reproducing results presented herein as is. Please see
this
\href{https://github.com/bondhugula/llvm-project/blob/hop/mlir/benchmark/README.md}{README}
to run most of the experiments.

The {\it hop} branch however is not intended to be continuously kept in sync
with changing upstream MLIR development --- so that the examples,
IR and results presented herein remain valid. On the other hand, the
\href{https://github.com/polymage-labs/mlirx}{MLIRX}~\cite{mlirx}
repository provides all of the features presented here while being maintained to
be in sync with upstream MLIR development. If the reader's objective is to build
upon the augmented IR infrastructure as opposed to just trying to reproduce
things, MLIRX is recommended.

\bibliographystyle{unsrt} \bibliography{references}

\end{document}